\documentclass[10pt]{article}
\usepackage{array}
\RequirePackage[OT1]{fontenc}
\RequirePackage{amsthm,amsmath,natbib}
\RequirePackage[colorlinks,citecolor=blue,urlcolor=blue]{hyperref}
\usepackage[psamsfonts]{amsfonts}
\usepackage{graphicx} 
\usepackage{lscape}

\numberwithin{equation}{section}
\theoremstyle{plain}
\newtheorem{Lem}{Lemma}[section]
\newtheorem{Prop}{Proposition}[section]

\def\Black{} 

\newcommand{\cqfd}{\hfill $\square$}

\newcommand{\R}{\mathbb R}

\newcommand{\N}{\mathbb N}

\newcommand{\n}{^{(n)}}

\newcommand{\Xb}{\mathbf{X}}
\newcommand{\Sb}{\mathbf{S}}

\newcommand{\vb}{\ensuremath{\mathbf{v}}}
\newcommand{\xb}{\ensuremath{\mathbf{x}}}

\newcommand{\Ub}{\ensuremath{\mathbf{U}}}

\newcommand{\Omegab}{{\pmb \Omega}}
\newcommand{\omegab}{{\pmb \omega}}

\newcommand{\thetab}{{\pmb \theta}}

\newcommand{\varthetab}{{\pmb \vartheta}}

\newcommand{\Thetab}{{\pmb \Theta}}

\newcommand{\Deltab}{{\pmb \Delta}}

\newcommand{\etab}{{\pmb \eta}}

\newcommand{\Gamb}{{\pmb \Gamma}}

\newcommand{\pr}{^{\prime}}

\newcommand{\uut}[1] {\renewcommand{\arraystretch}{0.5}
\begin{array}[t]{c}{#1}\vspace{.2mm}\\
   \widetilde{}\vspace{-2.1mm}
\end{array}
\renewcommand{\arraystretch}{1} }

\newcommand{\utDelta}{\!\! {\uut {\pmb \Delta}}\vspace{-1mm}\!\!}

\newcommand{\ny}{n\rightarrow\infty}

\newcommand{\lbar}{{\bar{\phantom u}}\hspace{-1.8mm}l}
\newcommand{\dbar}{{\bar{\phantom u}}\hspace{-2mm}d}

\begin{document}
\renewcommand{\baselinestretch}{1.2}
\markright{
}
\markboth{\hfill{\footnotesize\rm Ley et al.
}\hfill}
{\hfill {\footnotesize\rm Optimal $R$-estimation of a spherical location} \hfill}
\renewcommand{\thefootnote}{}
$\ $\par
\fontsize{10.95}{14pt plus.8pt minus .6pt}\selectfont
\vspace{0.8pc}
\centerline{\large\bf OPTIMAL \emph{R}-ESTIMATION OF A SPHERICAL LOCATION}
\vspace{.4cm}
\centerline{Christophe Ley$^{1}$, Yvik Swan$^{2}$, Baba Thiam$^{3}$ and Thomas Verdebout$^{3}$}
\vspace{.4cm}
\centerline{\it $^{1}$D\'epartement de Math\'ematique, E.C.A.R.E.S., Universit\'e Libre de Bruxelles, Belgique}
\centerline{\it $^{2}$UR en  Math\'ematique, Universit\'e du Luxembourg, Luxembourg}
\centerline{\it $^{3}$Laboratoire EQUIPPE, Universit\'e Lille Nord de France, France}
\vspace{.55cm}
\fontsize{9}{11.5pt plus.8pt minus .6pt}\selectfont

\begin{quotation}
\noindent {\it Abstract:}
In this paper, we provide $R$-estimators of the location of a rotationally
symmetric distribution on the unit sphere of $\R^k$. In order to do so we first
prove the local asymptotic normality property of a sequence of
rotationally symmetric models; this is a non standard result due to the
curved nature of the unit sphere. We then construct our estimators  by adapting
the Le Cam one-step methodology to spherical statistics and ranks. We show that
they are asymptotically normal under any rotationally symmetric
distribution and achieve the efficiency bound under a specific density.
Their small sample behavior is studied via a Monte Carlo simulation and
our methodology is illustrated on geological data.\par

\vspace{9pt}
\par
\end{quotation}\par

\fontsize{10.95}{14pt plus.8pt minus .6pt}\selectfont
\setcounter{equation}{0} 
\section{Introduction}\label{intro}

Spherical data arise naturally in a broad range of natural sciences such as  geology and astrophysics (see, e.g., \citet{Wat1983} or \citet{MarJup2000}), as well as in studies of animal behavior (see \citet{FisLewEmb87}) or  even in  neuroscience  (see \citet{LC98}). It is common practice to view such data as realizations of random vectors $\Xb$  taking values on the surface of the unit sphere $\mathcal{S}^{k-1}:= \{{\bf v} \in \R^k\,|\, \| {\bf v} \|=1\}$,
%
the  distribution of $\Xb$ depending only  on its distance  -- in a sense to be made precise~-- from a fixed point $\thetab \in \mathcal{S}^{k-1}$. This parameter $\thetab$,  which can be viewed as a  ``north pole'' (or ``mean direction'') for the problem under study,  is then a location parameter for the distribution.

The first   distribution tailored to the specificities of spherical data  is the {\em Fisher-Von Mises-Langevin} (FVML) distribution  introduced in   \citet{Fis1953}.  To this date it  remains the  distribution which is the most widely used in practice; it plays, in the spherical context, the central role enjoyed by the Gaussian distribution for linear data.  The FVML is, obviously,  not the only so-called ``spherical distribution'', and there exists  a wide variety of families of such  distributions possessing different advantages and drawbacks (see \citet{MarJup2000} for an overview).

In this work we will concentrate our attention on the family of \emph{rotationally symmetric distributions}  introduced by  \citet{Saw1978} (see Section \ref{sec:ULAN} below for a definition). 
Aside from the fact that it encompasses many well-known spherical distributions (including the FVML), this family satisfies a natural requirement~: it is invariant to  the actual choice of ``north pole''.  This entails that the family falls within the much more general class of \emph{statistical group models} (see for instance \citet{Cha2004}) and thus enjoys all the advantages of this class.
Moreover, it satisfies the following fundamental lemma, due to  \citet{Wat1983}.
\begin{Lem}[{\bf \citet{Wat1983}}]\label{wat} If the distribution of $\Xb$ is rotationally symmetric  on $\mathcal{S}^{k-1}$  and if the true location is  $\thetab$, then (i) $\Xb\pr \thetab$ and \linebreak $\Sb_\thetab(\Xb):={(\Xb- (\Xb\pr \thetab) \thetab)}/\|{\Xb- (\Xb\pr \thetab) \thetab}\|$ are stochastically independent and (ii)  the multivariate sign vector $\Sb_\thetab(\Xb)$ is uniformly distributed on ${\mathcal S}^{k-1}(\thetab^{\perp}):= \{ \vb \in \R^k\,|\, \| \vb \|=1, \vb\pr \thetab=0\}$.
\end{Lem}

Estimation and testing procedures for the spherical location parameter $\thetab$ have been extensively studied in the literature, with much of the focus in the past years being put on  the class of $M$-estimators. An $M$-estimator $\hat\thetab$ associated with a given function $\rho_0(\xb; \thetab)$ is defined  as the value of $\thetab$ which minimizes the objective function
$$\thetab\mapsto\rho(\thetab) := \sum_{i=1}^n \rho_0(\Xb_i; \thetab),$$ \Black
where $\Xb_1,\ldots,\Xb_n$ are spherical observations. These $M$-estimators are robust to outliers (see \cite{KoCha1993})  and enjoy nice asymptotic properties (see \citet{ChaRiv2001}, \citet{ChaTsa2003} or \citet{Cha2004}). In particular, the choice  $\rho_0(\xb; \thetab) = \arccos(\xb\pr\thetab)$ yields the so-called \emph{spherical median} introduced by   \citet{Fis1985}; taking    $\rho_0(\xb; \thetab) =\|\xb-\thetab\|^2$ yields $\hat\thetab = \bar{\Xb}/\|\bar{\Xb}\|$, the \emph{spherical mean}.

The first to have  studied ranks and rank-based methods within the spherical framework are  \citet{NeeCha2001},  who construct rank score statistics  of the form ${\bf T}_{\varphi; \thetab}\n:= \sum_{i=1}^n \varphi\left({R_i^+}/ ({n+1})\right) \Sb_\thetab(\Xb_i)$, \Black
where $\varphi$ is a score generating function and where $R_i^+$ ($i=1,\ldots,n$)  stands for the rank of $\|\Xb_i-(\Xb_i\pr\thetab)\thetab\|$ among  the scalars $\|\Xb_1-(\Xb_1\pr\thetab)\thetab\|, \ldots, \|\Xb_n-(\Xb_n\pr\thetab)\thetab\|$. 
Some years later,  \citet{TsaSen2007} developed rank tests (using a similar definition for the ranks $R_i^+$)  for the location problem $\mathcal{H}_{0}: \thetab= \thetab_0$. Assuming that the observations $\Xb_1, \ldots, \Xb_n$ are independent and identically distributed (\mbox{i.i.d.}) with a rotationally symmetric distribution on the unit sphere, 
they consider statistics of the form $({\bf T}_{\varphi; \thetab_0}\n)\pr \Gamb_{\bf T}^{-1}{\bf T}_{\varphi; \thetab_0}\n$, where $\Gamb_{\bf T}$ is the asymptotic variance of ${\bf T}_{\varphi; \thetab_0}\n$ under the null. They obtain the asymptotic properties of their procedures via permutational central limit theorems. 

The purpose of the present work is to propose optimal rank-based estimators ($R$-estimators) for the spherical location $\thetab$.  
The backbone of our approach is the so-called ``Le Cam methodology'' (see \citet{Lec1986}), which has, to the best of our knowledge, never been used in the framework of spherical statistics. This is perhaps explained by the curved nature of the parameter space: the unit sphere $\mathcal{S}^{k-1}$ being a non-linear manifold, it typically generates non-traditional Gaussian shift experiments and, as a consequence, the usual arguments behind Le Cam's theory break down in this context.

The first step consists in establishing the \emph{uniform local asymptotic normality} (ULAN) of a sequence of rotationally symmetric models. This we achieve by rewriting $\thetab\in\mathcal{S}^{k-1}$ in terms of the usual spherical $\etab$-coordinates and showing that ULAN holds for this more common re-parameterization. Although the latter parameterization is not valid uniformly on the whole sphere (its Jacobian is not full-rank everywhere), we then use a recent result of   \citet{HalPaiVer2010}  to show how the ULAN result in the $\etab$-parameterization carries through to the original $\thetab$-parameterization.

The second step consists in adapting to the spherical context the so-called method of {\it one-step $R$-estimation ``\`{a} la Le Cam''}, introduced in \citet{HalOjaPai2006} in the context of the estimation of a shape matrix of a multivariate elliptical distribution. The ULAN property mentioned above guarantees that the resulting rank-based estimators enjoy the desired optimality properties.

Our estimators constitute attractive  alternatives to the  traditional Hodges-Lehmann  rank-based estimators which  have,  in general,  two major drawbacks. First, they are defined through minimization of  a rank-based function which  is therefore non-continuous; this complicates greatly   the study of their asymptotic properties.  Secondly, as shown via a Monte Carlo study in \cite{HSVV2011} in a regression context, the performances of Hodges-Lehmann-type estimators tend to deteriorate as (the dimension) $k$ increases. As will be shown in this paper,  one-step $R$-estimators suffer from neither of these flaws and are even good competitors against the $M$-estimators studied in the literature.

The outline of the paper is as follows. In  Section \ref{sec:ULAN}  we recall the definition of the family of rotationally symmetric distributions and  prove  uniform local asymptotic normality of this family.  We devote Section \ref{sec:estimation} to a description of our rank-based estimators and to the study of their asymptotic properties.  In  Section \ref{sec:ARE}  we compare our $R$-estimators with the $M$-estimators from the literature in terms of Asymptotic Relative Efficiency as well as by means of a Monte Carlo study.  In Section~\ref{empap} we  apply our  $R$-estimators to geological data.  Finally an Appendix collects the technical proofs.

\section{Spherical model and ULAN}\label{sec:ULAN}
Throughout, the data points $\Xb_1, \ldots, \Xb_n$ are assumed to belong to the unit sphere $\mathcal{S}^{k-1}$  of $\R^k$ and to satisfy the following assumption.
\vspace{4mm}

{\sc Assumption A.}  $\Xb_1, \ldots, \Xb_n$ are \mbox{i.i.d.} with common distribution ${\rm P}_{\thetab; f_1}$ characterized by a density function (with respect to the usual surface area measure on spheres)
\begin{equation}\label{density}
\xb\mapsto f_{\thetab}({\xb})= c_{k,f_1} \; f_1({\bf x}\pr \thetab),\quad\xb\in\mathcal{S}^{k-1},\end{equation}
where $\thetab \in \mathcal{S}^{k-1}$ is a location parameter and $f_1: [-1,1] \rightarrow \R_0^+$  is  absolutely continuous and (strictly) monotone increasing. \vspace{4mm}

A function $f_1$ satisfying Assumption A will be called an \emph{angular function}; we throughout denote by $\mathcal{F}$ the set of angular functions. \Black This choice of terminology reflects the fact that, under Assumption A,   the distribution of $\Xb$ depends only on the  angle between it and  some location $\thetab\in\mathcal{S}^{k-1}$. If  $\Xb_1, \ldots, \Xb_n$ are \mbox{i.i.d.} with density (\ref{density}), then $\Xb_1\pr \thetab, \ldots, \Xb_n\pr \thetab$ are \mbox{i.i.d.} with density
$$t\mapsto\tilde{f}_1(t):= \frac{\omega_k \; c_{k,f_1}}{B(\frac{1}{2}, \frac{1}{2}(k-1))} f_1({t}) (1-t^2)^{(k-3)/2}, \quad -1\leq t \leq 1,$$
where $\omega_k= 2 \pi^{k/2}/ \Gamma (k/2)$ is the surface area of $\mathcal{S}^{k-1}$ and $B(\cdot,\cdot)$ is the beta function. The corresponding cdf will be denoted by $\tilde{F}_1(t)$.

A special instance of (\ref{density}) is the FVML  distribution, obtained by taking {angular functions} of the form  $f_1(t)= {\rm exp}(\kappa t)=:f_{1; {\rm exp}}(\kappa t)$ for some concentration parameter $\kappa>0$. 
Aside from the FVML distribution we will also consider spherical distributions with  angular functions
\begin{equation}\label{linlog} f_{1; {\rm Lin(a)}}(t):=t+a,\quad  f_{1; {\rm Log(a)}}(t):=\log(t+a) \quad \mbox{  and } \end{equation} \begin{equation} \label{densities} f_{1; {\rm Logis}(a,b)}(t):= \frac{a \; {\rm exp}(-b \; {\rm arccos}(t))}{(1+a \; {\rm exp}(-b \; {\rm arccos}(t)))^2}; \end{equation}
we refer to these as the  \emph{linear}, \emph{logarithmic} and \emph{logistic} spherical distributions, respectively ($a$ and $b$ are constants chosen so  that all the above angular functions belong to $\mathcal{F}$). 


The rest of this section is devoted to the establishment of the ULAN property of the family $\{{\rm P}^{(n)}_{\thetab; f_1} \; \vert \; \thetab \in \mathcal{S}^{k-1}\}$, where ${\rm P}^{(n)}_{\thetab; f_1}$ stands for the joint distribution of $\Xb_1,\ldots,\Xb_n$;   see the comment just below Proposition \ref{ULAN} for a definition of ULAN.   As mentioned in the Introduction, in order to obtain this ULAN property we  first need to circumvent a difficulty inherited from the curved nature of the experiment we are considering.

Among the first to have considered such ``curved experiments'',    \citet{ChaRiv2001} and \citet{Cha2004} suggest to bypass the problem by   reformulating  the notion of Fisher information in terms of
 inner products on the tangent space to $\mathcal{S}^{k-1}$. 
 In this paper, we rather adopt an approach based on recent results from  \citet{HalPaiVer2010} and, in particular, on the following lemma.
\Black
\begin{Lem}[{\bf \citet{HalPaiVer2010}}]\label{LElemme}
Consider a family of probability distributions ${\cal P}\n = \{{\rm P}\n_\omegab\ \vert \ \omegab\in{\pmb\Omega}\}$ with $\Omegab$ an open subset of $\R^{k_1}$ ($k_1\in\N_0$). Suppose that the parameterization $\omegab\mapsto{\rm P}\n_\omegab$ is ULAN for ${\cal P}\n$ at some point $\omegab_0\in\Omega$, with central sequence $\Deltab\n_{\omegab_0}$ and Fisher information matrix $\Gamb_{\omegab_0}$. Let $\dbar : \omegab\mapsto \varthetab := \dbar(\omegab)$ be a continuously differentiable mapping  from $\R^{k_1}$ to  $\R^{k_2}$ ($k_1\leq k_2\in\N_0$) with full column rank Jacobian matrix $D\dbar (\omegab)$ at every $\omegab$ in some neighborhood of $\omegab_0$. Write $\Thetab := \dbar(\Omegab )$, and assume that $\varthetab\mapsto {\rm P}^{(n);\,\dbar}_\varthetab$,  $\varthetab\in{\Thetab}$, provides another parameterization of ${\cal P}\n$.
Then $\varthetab\mapsto {\rm P}^{(n);\,\dbar}_\varthetab$ is also ULAN for ${\cal P}\n$ at $\varthetab_0=\dbar (\omegab_0)$, with central sequence $\Deltab^{(n);\,\dbar}_{\varthetab_0} = (D^-\dbar (\omegab_0 ) )\pr\Deltab^{(n)}_{\omegab_0}$  and Fisher information matrix $\Gamb^{\dbar}_{\varthetab_0}   = (D^-\dbar (\omegab_0 ))\pr\Gamb_{\omegab_0} D^-\dbar (\omegab_0 )$, where $D^-\dbar (\omegab_0 ):=((D\dbar (\omegab_0 ) )\pr D\dbar (\omegab_0 ))^{-1}(D\dbar (\omegab_0 ))\pr $ is the Moore-Penrose inverse of~$D\dbar (\omegab_0 )$.
\end{Lem}



We start by establishing ULAN for a re-parameterization of the problem in terms of spherical coordinates.  Any vector $\thetab$ on the unit sphere of $\R^k$ can be represented via the chart
\begin{align}\label{param1}
\hbar: \etab:= &  \left( \eta_{1}, \ldots, \eta_{k-1}\right)\pr \in \R^{k-1} \mapsto \hbar(\etab)= \thetab= ( \cos \eta_1, \sin \eta_1 \; \cos \eta_2, \\
		       &  \quad \quad \quad \quad  \quad \quad \ldots, \sin \eta_1 \cdots \sin \eta_{k-2} \; \cos \eta_{k-1}, \sin \eta_1 \cdots \sin \eta_{k-2} \; \sin \eta_{k-1})\pr,\nonumber 
\end{align}
%
whose Jacobian matrix $D\hbar(\etab)$ is given by
\begin{equation*}
\left(
\begin{array}{ccccc}
-\sin \eta_1&0&\ldots&0\\
\cos \eta_1\cos\eta_2&-\sin\eta_1\sin\eta_2&\ldots&0\\
\vdots&\vdots&\vdots&\vdots\\
\cos\eta_1\prod_{j=2}^{k-3}\sin\eta_j\cos \eta_{k-2}&\sin\eta_1\cos\eta_2\prod_{j=3}^{k-3}\sin \eta_j \cos\eta_{k-2}&\ldots &0\\
\cos\eta_1\prod_{j=2}^{k-2}\sin\eta_j\cos \eta_{k-1}&\sin\eta_1\cos\eta_2\prod_{j=3}^{k-2}\sin \eta_j \cos \eta_{k-1}&\ldots&-\prod_{j=1}^{k-1}\sin\eta_j\\
\cos\eta_1\prod_{j=2}^{k-1}\sin\eta_j&\sin\eta_1\cos\eta_2\prod_{j=3}^{k-1}\sin \eta_j&\ldots&\prod_{j=1}^{k-2}\sin\eta_j\cos\eta_{k-1}
\end{array}
\right).
\end{equation*}
%
%
The corresponding ${\pmb \eta}$-parameterization $\{ {\rm P}_{\etab; f_1}^{(n);\hbar} \;  \vert \; {\pmb \eta} \in \R^{k-1}\}$ is simple to construct and enjoys the advantage of a classical parameter space, namely $\R^{k-1}$. This  is very convenient for asymptotic calculations; in particular, it is helpful when quadratic mean differentiability and consequently ULAN have to be proved. 

Before proceeding to the statement of our results  we need the following (essentially technical) assumption.

\

{\sc Assumption B}. Letting $\varphi_{f_1}:=\dot{f_1}/f_1$ ($\dot{f_1}$ is the a.e.-derivative of $f_1$), the quantity $\mathcal{J}_k({f_1}):= \int_{-1}^{1} \varphi_{f_1}^2(t) (1-t^2) \tilde{f}_1(t) dt < + \infty$.

\

\noindent Assumption B entails that the Fisher information matrix for spherical location is finite  (in both the $\etab$- and the original $\thetab$-parameterization). More precisely, we will show that the information matrix for the $\etab$-parameterization with chart $\hbar$ is of the form  $$\frac{\mathcal{J}_k({f_1})}{k-1}D\hbar(\etab) \pr ({\bf I}_k- \hbar(\etab)\hbar(\etab)\pr) D\hbar(\etab)$$ with ${\bf I}_k$ standing for the $k\times k$ identity matrix.
This matrix can be rewritten as $\frac{\mathcal{J}_k({f_1})}{k-1}{\pmb \Omega}_{k,\etab}^{\hbar}$, where
\begin{eqnarray*}{\pmb \Omega}_{k,\etab}^{\hbar}&:= & {\rm diag}(1, \sin^2 \eta_1, \sin^2 \eta_1 \sin^2 \eta_2, \ldots, \sin^2 \eta_1 \cdots \sin^2 \eta_{k-2})\\ &=& D\hbar(\etab)\pr D\hbar(\etab) \\ & =&D\hbar(\etab) \pr ({\bf I}_k- \hbar(\etab)\hbar(\etab)\pr) D\hbar(\etab), \end{eqnarray*}
with ${\rm diag}(a_1, \ldots, a_l)$ denoting a diagonal matrix with diagonal elements $a_1, \ldots, a_l$. Hence, if $ D\hbar(\etab)$ is full column rank, then the information matrix  ${\pmb \Omega}_{k,\etab}^{\hbar}$ is also full-rank.

With this in hand we are able to establish the ULAN property of the family $\{ {\rm P}_{{\pmb \eta}; f_1}^{(n);\hbar} \;  \vert \; {\pmb \eta} \in \R^{k-1}\}$ at any point $\etab_0\in\R^{k-1}$ (note that, clearly, at some points the information matrix will be singular, due to the rank deficiency of the Jacobian matrix). In order to avoid heavy notations, we drop the index 0 in what follows and write out the LAN property at $\etab\in\R^{k-1}$ in the following proposition (see the Appendix for a proof).


\begin{Prop}\label{ULAN} Let Assumptions A and B hold. Then the family of probability distributions $\left\{ {\rm P}_{{\pmb \eta}; f_1}^{(n);\hbar} \;  \vert \; {\pmb \eta} \in \R^{k-1} \right\}$ is LAN with central sequence
\begin{equation}\label{centrseq}\Deltab_{{\pmb \eta}; f_1}^{(n);\hbar}:= n^{-1/2} \sum_{i=1}^{n} \varphi_{f_1}(\Xb_i \pr \hbar(\etab)) (1-(\Xb_i \pr \hbar(\etab))^2)^{1/2}  D\hbar(\etab) \pr \Sb_{\hbar(\etab)}(\Xb_i)\end{equation}
and Fisher information matrix
\begin{equation} \label{FishF} \Gamb_{\etab; f_1}^{\hbar}:= \frac{\mathcal{J}_k({f_1})}{k-1} {\pmb \Omega}_{k,\etab}^{\hbar}.  \end{equation}
More precisely, for any bounded sequence ${\bf e}\n \in \R^{k-1}$,
\begin{equation} \label{quad}
\log \left( \frac{d{\rm P}_{{\pmb \eta} + n^{-1/2} {\bf e}\n; f_1}^{(n);\hbar}}{d{\rm P}_{{\pmb \eta}; f_1}^{(n);\hbar}}\right)=  ({\bf e}\n)\pr \Deltab_{{\pmb \eta}; f_1}^{(n);\hbar} - \frac{1}{2}({\bf e}\n)\pr \Gamb_{ \etab; f_1}^{\hbar} {\bf e}\n +o_{\rm P}(1) \end{equation}
and $\Deltab_{{\pmb \eta}; f_1}^{(n);\hbar}\stackrel{\mathcal{L}}{\rightarrow}\mathcal{N}_{k-1}({\bf 0},\Gamb_{\etab; f_1}^{\hbar})$, both under ${\rm P}_{{\pmb \eta}; f_1}^{(n);\hbar}$, as $\ny$.
\end{Prop}

In the sequel, we will make use of a slightly reinforced version of Proposition~\ref{ULAN}, namely the ULAN property of our sequence of models. 
A sequence of rotationally symmetric models is called ULAN if, for any $\etab\n$ such that $\etab\n-\etab=O(n^{-1/2})$ and any bounded sequence ${\bf e}\n \in \R^{k-1}$,
$$
\log \left( \frac{d{\rm P}_{{\pmb \eta}\n + n^{-1/2} {\bf e}\n; f_1}^{(n);\hbar}}{d{\rm P}_{{\pmb \eta}\n; f_1}^{(n);\hbar}}\right)=  ({\bf e}\n)\pr \Deltab_{{\pmb \eta}\n; f_1}^{(n);\hbar} - \frac{1}{2}({\bf e}\n)\pr \Gamb_{ \etab; f_1}^{\hbar} {\bf e}\n +o_{\rm P}(1)
$$
and $\Deltab_{{\pmb \eta}\n; f_1}^{(n);\hbar}\stackrel{\mathcal{L}}{\rightarrow}\mathcal{N}_{k-1}({\bf 0},\Gamb_{\etab; f_1}^{\hbar})$, both under ${\rm P}_{{\pmb \eta}\n; f_1}^{(n);\hbar}$, as $\ny$. Note that the information matrix $\Gamb_{ \etab; f_1}^{\hbar}$ is uniformly (in $\etab$) the same in the neighborhood $\left\{\etab\n, \etab\n-\etab=O(n^{-1/2})\right\}$ of $\etab$. In the present setup, it can easily be verified that ULAN holds by using the same arguments as in the proof of Proposition~\ref{ULAN}.

Note that the diagonality of the Fisher information \eqref{FishF} is convenient when working in the $\etab$-parameterization. Indeed it  is the structural reason why, when performing asymptotic inference on $\eta_1$ for example,  the non-specification of the parameters $\eta_2, \ldots, \eta_{k-1}$ will not be responsible for any loss of efficiency. Our focus being on the $\thetab$-parameterization, we will not  investigate this issue any further.


The $\etab$-parameterization obtained via the chart $\hbar$ suffers from the drawback that the information matrix is singular at several points on the sphere. More precisely, if $\sin \eta_l=0$ for some $\eta_l \in \{\eta_1, \ldots, \eta_{k-2}\}$, then ${\pmb \Omega}_{k,\etab}^{\hbar}$ is singular. For example, in the $3$-dimensional case, the parameter values $\thetab=(1, {\bf 0}\pr)\pr$ or $\thetab=(-1, {\bf 0}\pr)\pr$ are very particular; for those values,  ${\pmb \Omega}_{3,\etab}^{\hbar}={\rm diag}(1,0)$. This phenomenon is due to an identification problem: in the general $k$-dimensional case, when $\eta_1=0$ for example, then ${\rm P}_{\eta_1, \eta_2, \ldots, \eta_{k-1}; f_1}={\rm P}_{\eta_1, \tilde\eta_2, \ldots, \tilde\eta_{k-1}; f_1}$ for any $(k-2)$-uples $(\eta_2, \ldots, \eta_{k-1}) \neq (\tilde\eta_2, \ldots, \tilde\eta_{k-1})$. In the same way, whenever $\eta_j=0$ for some $j \in \{1, \ldots, k-2 \}$, the Jacobian matrix $D\hbar(\etab)$ is not full-rank and as a consequence $\Gamb_{\etab; f_1}^{\hbar}$ is singular. 
This singularity -- and the identification problem from which it results --  is however not structural and only originates in the choice of the chart.

  Since every point of a $m$-dimensional manifold has a neighborhood homeomorphic to an open subset of the $m$-dimensional space $\R^m$, we see that, for all $\thetab \in {\mathcal S}^{k-1}$, one can find a chart $\lbar: \R^{k-1} \mapsto \mathcal{S}^{k-1}:\etab\mapsto \thetab=\lbar(\etab)$ with a full column rank Jacobian matrix $D\lbar(\etab)$ in the vicinity of $\etab$. Making use of Proposition~\ref{ULAN}, whose proof does not involve the particular form of $\hbar$, we obtain that the family $\{ {\rm P}_{{\pmb \eta}; f_1}^{(n);\,\lbar} \;  \vert \; {\pmb \eta} \in \R^{k-1}\}$ is ULAN with central sequence $$\Deltab_{{\pmb \eta}; f_1}^{(n);\,\lbar}:= n^{-1/2} \sum_{i=1}^{n} \varphi_{f_1}(\Xb_i \pr \lbar(\etab)) (1-(\Xb_i \pr \lbar(\etab))^2)^{1/2}  D\lbar(\etab) \pr \Sb_{\lbar(\etab)}(\Xb_i)$$ and full-rank Fisher information matrix $$ \Gamb_{\etab; f_1}^{\lbar}=\frac{\mathcal{J}_k({f_1})}{k-1} {\pmb \Omega}_{k,\etab}^{\lbar}=\frac{\mathcal{J}_k({f_1})}{k-1} D\lbar(\etab)\pr D\lbar(\etab).$$
This shows that, for any $\thetab\in\mathcal{S}^{k-1}$, it is possible to find a chart $\lbar$ such that,  in a neighborhood of $\etab=\lbar^{-1}(\thetab)$, the Jacobian matrix is non-singular and the ULAN property for the related family $\left\{ {\rm P}_{{\pmb \eta}; f_1}^{(n);\,\lbar} \;  \vert \; {\pmb \eta} \in \R^{k-1} \right\}$ holds. This observation, combined with Lemma~\ref{LElemme}, finally yields the desired ULAN property for the family $\left\{ {\rm P}_{\thetab; f_1}\n \;  \vert \; {\thetab} \in {\mathcal S}^{k-1} \right\}$ (see the Appendix for a proof).

Let $\thetab\n \in {\mathcal S}^{k-1}$ be such that $\thetab\n-\thetab=O(n^{-1/2})$ and consider local alternatives on the sphere of the form  $\thetab\n + n^{-1/2}{\bf t}\n$. For ${\thetab}\n+ n^{-1/2}{\bf t}\n$ to remain in ${\mathcal S}^{k-1}$, it is necessary that  the sequence ${\bf t}\n$  satisfies \begin{eqnarray} \label{localt} 0 &=& ({\thetab}\n+ n^{-1/2}{\bf t}\n)\pr ({\thetab}\n+ n^{-1/2}{\bf t}\n) -1 \nonumber \\ &=&  2n^{-1/2}({\thetab}\n)\pr {\bf t}\n + n^{-1} ({\bf t}\n)\pr {\bf t}\n. \end{eqnarray}
Consequently, ${\bf t}\n$ must be such that $2n^{-1/2}\; ({\thetab}\n)\pr  {\bf t}\n + n^{-1} ({\bf t}\n)\pr {\bf t}\n=0$ or equivalently such that $2 n^{-1/2}({\thetab}\n)\pr  {\bf t}\n+ o(n^{-1/2})=0$. Therefore, for ${\thetab}\n+ n^{-1/2}{\bf t}\n$ to remain in ${\mathcal S}^{k-1}$, ${\bf t}\n$ must belong, up to a $o(n^{-1/2})$ quantity, to the tangent space to $\mathcal{S}^{k-1}$ at $\thetab\n$.


Now, Lemma~\ref{LElemme} provides a link between the ULAN properties of two different parameterizations of the same model. This link is directly related to the link between the local alternatives $\etab\n+n^{-1/2}{\bf e}\n$ in the full-rank parameterization and the local alternatives $\thetab\n + n^{-1/2}{\bf t}\n$ described above. As shown in (\ref{localt}), in order for $\thetab\n + n^{-1/2}{\bf t}\n$ to belong to $\mathcal{S}^{k-1}$, it is necessary that ${\bf t}\n$ be of the form ${\bf t}_*\n + o(1)$, with ${\bf t}_*\n$ in the tangent space  to $\mathcal{S}^{k-1}$ at $\thetab\n$, hence of the form  ${\bf t}_{**}\n + o(1)$ with ${\bf t}_{**}\n$ in the tangent space  to $\mathcal{S}^{k-1}$ at $\thetab$ that is,  ${\bf t}_{**}\n=D\lbar (\etab){\bf e}\n$ for some bounded sequence ${\bf e}\n\in\R^{k-1}$.  It follows from differentiability that, letting  $\etab\n=\lbar^{-1}(\thetab\n )$, 
\begin{eqnarray}\nonumber 
\thetab\n\! + n^{-1/2}{\bf t}\n\! \!&\! =\!& \! \thetab\n + n^{-1/2}D\lbar (\etab){\bf e}\n + o(n^{-1/2})
\\
\! &\! =\!& \!  
 \lbar(\etab\n) + n^{-1/2}D\lbar (\etab){\bf e}\n  + o(n^{-1/2}) \label{c'estlui} \nonumber \\
\!& \! = \!& \!
\lbar(\etab\n + n^{-1/2}  {\bf e}\n + o(n^{-1/2})),
\end{eqnarray}
hence there is a clear correspondence between linear perturbations  in the $\etab$-parameterization and perturbations on the sphere in the $\thetab$-parameterization (through the chart $\lbar$). As a direct consequence, the local alternatives $\thetab\n + n^{-1/2}{\bf t}\n$ are equivalent to those considered in \cite{Tsa2009}.
Now, turning to local log-likelihood ratios, in view of ULAN for the $\etab$-parameterization,
\begin{eqnarray*}\nonumber
\log\Big(d{\rm P}\n_{\thetab\n + n^{-1/2}{\bf t}\n; f_1}/d{\rm P}\n_{\thetab\n; f_1 }\Big)  
&=&
\log\Big(d{\rm P}\n_{ \etab\n + n^{-1/2}   {\bf e}\n + o(n^{-1/2}); f_1}/d{\rm P}\n_{\etab\n; f_1}\Big) \nonumber\\
&=& {\bf e}^{(n)\prime}\Deltab_{{\pmb \eta}\n; f_1}^{(n);\,\lbar} -\frac{1}{2} \, {\bf e}^{(n)\prime} \Gamb_{\etab; f_1}^{\lbar}  {\bf e}\n +o_{\rm P}(1)
\end{eqnarray*}
under 
${\rm P}\n_{ \etab\n; f_1} = {\rm P}\n_{ \varthetab\n; f_1}$-probability, 
as $\ny$. Summing up, we have the following result.


\begin{Prop}\label{ULANbis} Let Assumptions A and B hold. Then the family of probability distributions $\left\{ {\rm P}_{\thetab; f_1}^{(n)} \;  \vert \; \thetab \in \mathcal{S}^{k-1} \right\}$ is ULAN with central sequence
\begin{equation*}
\Deltab_{\thetab; f_1}\n:=n^{-1/2} \sum_{i=1}^{n} \varphi_{f_1}(\Xb_i \pr \thetab) (1-(\Xb_i \pr \thetab)^2)^{1/2}  \Sb_\thetab(\Xb_i)
\end{equation*}
and Fisher information matrix
\begin{equation*}
\Gamb_{\thetab; f_1}:=\frac{\mathcal{J}_k({f_1})}{k-1}({\bf I}_k- \thetab \thetab\pr).
\end{equation*}
More precisely, for any $\thetab\n \in {\mathcal S}^{k-1}$ such that $\thetab\n-\thetab=O(n^{-1/2})$ and any bounded sequence ${\bf t}\n$ as in (\ref{localt}),
we have
$$\log \left( \frac{d{\rm P}_{\thetab\n + n^{-1/2} {\bf t}\n; f_1}^{(n)}}{d{\rm P}_{\thetab\n; f_1}^{(n)}}\right)=  ({\bf t}\n)\pr \Deltab_{\thetab\n; f_1}\n - \frac{1}{2}({\bf t}\n)\pr \Gamb_{\thetab; f_1} {\bf t}\n +o_{\rm P}(1)$$
and $\Deltab_{\thetab\n; f_1}\n\stackrel{\mathcal{L}}{\rightarrow}\mathcal{N}_{k-1}({\bf 0},\Gamb_{\thetab; f_1})$, both under ${\rm P}_{\thetab\n; f_1}^{(n)}$, as $\ny$.
\end{Prop}

See the proof in the appendix in which the expressions of $\Deltab_{{\thetab}; f_1}\n$ and $\Gamb_{\thetab; f_1}$, are obtained.

\Black

\setcounter{equation}{0} 
 \section{Rank-based estimation: optimal $R$-estimators}\label{sec:estimation}

In this section, we make use of the ULAN property of Proposition~\ref{ULANbis}  to construct semi-parametrically efficient $R$-estimators of $\thetab$. To this end, we adapt the Le Cam technique of one-step $R$-estimation, introduced in \citet{HalOjaPai2006}, to the present context.

Le Cam's  one-step  $R$-estimation method assumes the existence of a preliminary estimator  $\hat{\thetab}$ of $\thetab$ satisfying some conditions, which are summarized in the following assumption.

\

{\sc Assumption} C. The preliminary estimator  $\hat\thetab \in {\mathcal S}^{k-1}$ \Black is such that
\begin{itemize}
\item[(i)] $\hat{\thetab}- \thetab= O_{\rm p}(n^{-1/2})$ under $\bigcup_{g_1 \in \mathcal{F}} {\rm P}\n_{\thetab;g_1}$.
\item[(ii)] $\hat{\thetab}$ is \emph{locally and asymptotically discrete}; that is, it only takes a bounded number of distinct values in $\thetab$-centered balls with $O(n^{-1/2})$ radius. \vspace{2mm}
\end{itemize}

\

Assumption C(i) requires that the preliminary estimator is root-$n$ consistent under the whole set $\mathcal{F}$ of angular functions. This condition is fulfilled by, e.g., the spherical mean or the spherical median. This uniformity in $g_1 \in {\mathcal F}$ plays an important role for our $R$-estimation procedures, as we shall see in the sequel. Regarding Assumption C(ii), it should be noted that this discretization condition is a purely technical requirement (see, e.g., \citet{HSVV2011}), with little practical implications (in fixed-$n$ practice, such discretizations are irrelevant as the discretization radius can be taken arbitrarily large). Therefore, for the sake of simplicity, we tacitly assume in the sequel that $\hat{\thetab}$ satisfies Assumption C(ii).


\subsection{Rank-based central sequence and its asymptotic properties}\label{sec:est1}

The main idea behind one-step $R$-estimation consists in adding to the preliminary estimator $\hat\thetab$ a rank-based quantity which provides optimality under a fixed density. This rank-based quantity will  appear through a rank-based version $\utDelta_{{\thetab}; K}\n$ (rigorously defined in (\ref{rcentralseq}) below) of the parametric central sequence obtained in Proposition \ref{ULANbis}. 
A natural requirement for our rank-based central sequence~$\utDelta_{{\thetab}; K}\n$ is its distribution-freeness under the broadest possible family of distributions, say ${\rm P}\n_{\rm broad}$. \Black A classical way to achieve this goal consists in having recourse to the so-called \emph{invariance principle} which, when ${\rm P}\n_{\rm broad}$ is invariant under a group of transformations,  recommends expressing $\utDelta_{{\thetab}; K}\n$ in terms of  the corresponding maximal invariant. This entails that if, furthermore,  this group of transformations is a generating group for ${\rm P}\n_{\rm broad}$, then $\utDelta_{{\thetab}; K}\n$ is distribution-free under ${\rm P}\n_{\rm broad}$.

Clearly, in this rotationally symmetric context, invariance with respect to rotations is crucial, and therefore it seems at first sight natural to study ${\rm P}\n_{\rm broad}=\bigcup_{\thetab \in {\mathcal S}^{k-1}}{\rm P}\n_{\thetab; g_1}$ for fixed $g_1\in\mathcal{F}$ under the effect of the group of rotations. This group is generating for $\bigcup_{\thetab \in {\mathcal S}^{k-1}}{\rm P}\n_{\thetab; g_1}$, and the ranks $R_i^+$ defined in the Introduction are rotationally invariant because the scalar products $\Xb_{i}\pr \thetab$ are invariant with respect to rotations. This is not the case of the multivariate signs ${\Sb}_{\thetab} (\Xb_i)$, which are only rotationally equivariant in the sense that ${\Sb}_{\thetab} ({\bf O}\Xb_i)={\bf O}{\Sb}_{\thetab} (\Xb_i)$ for any ${\bf O}\in {\mathcal SO}_k$, the class of all $k\times k$ orthogonal matrices. However, since our aim consists in estimating $\thetab$, we rather work under fixed-$\thetab$ assumptions, hence the above-mentioned family ${\rm P}\n_{\rm broad}$ makes little sense in the present context. Furthermore, when inference on $\thetab$ is considered, the angular function $g_1$ remains an infinite-dimensional nuisance, which indicates that statistics which are invariant and therefore distribution-free under $\bigcup_{g_1 \in {\mathcal F}}{\rm P}\n_{\thetab; g_1}$ ($\thetab$ fixed) should in fact be taken into account, if possible without losing invariance properties with respect to the group of rotations.\Black

Fix $\thetab\in\mathcal{S}^{k-1}$ and consider ${\rm P}\n_{\rm broad}=\bigcup_{g_1 \in {\mathcal F}}{\rm P}\n_{\thetab; g_1}$. We obviously have that $\Xb_i = (\Xb_i\pr \thetab)  \thetab + \sqrt{1- (\Xb_i\pr\thetab)^2} {\Sb}_{\thetab} (\Xb_i)$ by definition of the multivariate signs. Now, let  $\mathcal{G}_h\n$ be the group of transformations of the form $g_h\n: (\Xb_1,\ldots,\Xb_n)  \mapsto(g_h(\Xb_1),\ldots,g_h(\Xb_n))$ with
\begin{equation}\label{monotrans}
g_h(\Xb_i):=h(\Xb_i\pr \thetab) \thetab + \sqrt{1- h(\Xb_i\pr\thetab)^2} {\Sb}_{\thetab} (\Xb_i),\quad i=1,\ldots,n,
\end{equation}
where $h:[-1,1]\rightarrow [-1,1]$ is a monotone continuous nondecreasing function such that $h(1)=1$ and $h(-1)=-1$. For any $g_h\n \in \mathcal{G}_h\n$, it is easy to verify that $\| {g}_h\n(\Xb_i) \|=1$; this means that $g_h\n \in \mathcal{G}_h\n$ is a monotone transformation from $\left({\mathcal S}^{k-1}\right)^n$ to $\left({\mathcal S}^{k-1}\right)^n$. It is quite straightforward to see that the group $\mathcal{G}_h\n$ is a generating group of the family of distributions $\bigcup_{f_1\in \mathcal F} {\rm P}_{\thetab; f_1}\n$. \Black

These considerations  naturally raise  the question of finding the maximal invariant associated with $\mathcal{G}_h\n$. Note that the definition of the mapping $g_h$ in (\ref{monotrans}), together with the fact that $\thetab\pr {\Sb}_{\thetab} (\Xb_i)=0$, entails that
\begin{eqnarray*}
g_h(\Xb_i)-(g_h(\Xb_i)\pr \thetab) \thetab &=& \sqrt{1- h(\Xb_i\pr\thetab)^2} {\Sb}_{\thetab} (\Xb_i).
\end{eqnarray*}
Then, we readily obtain that, for $i=1, \ldots,n$,
\begin{eqnarray*}
{\Sb}_{\thetab} (g_h(\Xb_i)) &=& \frac{g_h(\Xb_i)-(g_h(\Xb_i)\pr \thetab) \thetab}{\| g_h(\Xb_i)-(g_h(\Xb_i)\pr \thetab) \thetab\|} \\ &=& \frac{{\Sb}_{\thetab} (\Xb_i)}{\|Ê{\Sb}_{\thetab} (\Xb_i) \|}={\Sb}_{\thetab} (\Xb_i).
\end{eqnarray*}
This shows that the vector of signs ${\Sb}_{\thetab} (\Xb_1), \ldots, {\Sb}_{\thetab} (\Xb_n)$ is invariant under the action of $ \mathcal{G}_h\n$. However, it is not a maximal invariant; indeed, the latter is in most semiparametric setups composed of signs \emph{and} ranks (see for instance \cite{HalPai2006}).  Now, the ranks $R_i^+$ are not invariant under the action of $ \mathcal{G}_h\n$:  one can easily find a monotone transformation $g_{\tilde{h}}$ as in (\ref{monotrans}) such that the vector of ranks $R_i^+$ computed from $\Xb_1, \ldots, \Xb_n$ differs from the vector of ranks $R_i^+$ computed from  $g_{\tilde{h}}(\Xb_1), \ldots, g_{\tilde{h}}(\Xb_n)$. Thus, we need a different concept of ranks in order to build a rank-based version of our central sequence. For this purpose, define, for all $i=1,\ldots,n$, $R_i$ as the rank of $\Xb_i\pr\thetab$ among $\Xb_1\pr\thetab, \ldots, \Xb_n\pr\thetab$. 
Since in general, the ranks are invariant with respect to monotone transformations, noting that $g_h(\Xb_i)\pr \thetab=h(\Xb_i\pr \thetab)$ directly entails the invariance
of these new ranks  under the action of the group $\mathcal{G}_h\n$. The maximal invariant associated with $\mathcal{G}_h\n$ is, therefore, the vector of signs ${\bf S}_{\thetab}(\Xb_1), \ldots,  {\bf S}_{\thetab}(\Xb_n)$ and ranks $R_1, \ldots, R_n$. \Black

While it is preferable to use  the ranks $R_i$ when invariance with respect to $\mathcal{G}_h\n$ is required, there exist situations in which the ranks $R_i^+$ are appealing, e.g. when the angular functions are of the form $f_1(t)=\exp(g(t))$ with $g(-t)=-g(t)$ (since then the score function $\varphi_{f_1}(t)=\dot{g}(t)$ in (\ref{centrseq}) is symmetric), see \citet{TsaSen2007}.


As explained above, it follows that any statistic measurable with respect to the signs ${\bf S}_{\thetab}(\Xb_i)$ and ranks $R_i$ is distribution-free under $\bigcup_{g_1\in \mathcal F} {\rm P}_{\thetab; g_1}\n$. In accordance with these findings, we choose to base our inference  procedures  on the following sign- and rank-based version of the parametric central sequence obtained in Proposition \ref{ULANbis}:
\begin{equation} \label{rcentralseq} \utDelta_{\thetab; K}\n:=  n^{-1/2} \sum_{i=1}^{n} K \left(\frac{R_i}{n+1} \right) \Sb_\thetab(\Xb_i), \end{equation}
where $K$ is a score function satisfying

\

{\sc Assumption D}. The score function $K$ is a continuous function from $[0,1]$ to $\R$.

\

\noindent Note that  all the score functions associated with  the densities in (\ref{densities}) satisfy this assumption.

In the following result (the proof is given in the Appendix), we derive the asymptotic properties of $ \utDelta_{ \thetab; K}\n$ under ${\rm P}_{{\thetab}; g_1}\n$ and under contiguous alternatives ${\rm P}_{{\thetab}+ n^{-1/2}{\bf t}\n; g_1}\n$ (where ${\bf t}\n$ is a bounded sequence as described in $(\ref{localt})$)  for some angular function $g_1\in\mathcal{F}$.

\begin{Prop}  \label{asymplin}
Let Assumptions A, B and D  hold. Then the rank-based central sequence $\utDelta_{{\thetab}; K}\n$
\begin{itemize}
\item[(i)] is such that $\utDelta_{{\thetab}; K}\n-\Deltab_{{\thetab}; K, g_1}\n=o_{\rm P}(1)$ under ${\rm P}_{{\thetab}; g_1}\n$ as $\ny$, where ($\tilde{G}_1$ \vspace{0mm}

 stands for the common cdf of the $\Xb_i\pr \thetab$'s under ${\rm P}_{{\thetab}; g_1}\n$)
\begin{equation*}
\Deltab_{{\thetab}; K, g_1}\n:= n^{-1/2} \sum_{i=1}^{n} K_{}\left(\tilde{G}_1(\Xb_i\pr \thetab)\right)  \Sb_\thetab(\Xb_i).
\end{equation*}
Hence, for $K(u)= K_{f_1}(u)=\varphi_{f_1}(\tilde{F}_1^{-1}(u)) (1-(\tilde{F}_1^{-1}(u))^2)^{1/2}$, $\utDelta_{{\thetab}; K}\n$ is asymptotically equivalent to \vspace{0mm} the efficient  central sequence $\Deltab_{{\thetab}; f_1}\n$ for  spherical location under ${\rm P}_{\thetab; f_1}\n$.
\item[(ii)] is asymptotically normal under ${\rm P}_{{\thetab}; g_1}\n$ with mean zero and covariance matrix $\Gamb_{{\thetab};K}:=\frac{ {\mathcal J}_k({K})}{k-1} ({\bf I}_k- \thetab \thetab\pr)$ \Black, where ${\mathcal J}_k({K}):= \int_0^1 K^2(u) du$.

\item[(iii)] is asymptotically normal under ${\rm P}_{{\thetab}+n^{-1/2} {\bf t}\n; g_1}\n$ with mean $\Gamb_{{\thetab}; K, g_1} {\bf t}$ (where ${\bf t}:= \lim_{\ny} {\bf t}\n$) \Black and covariance matrix $\Gamb_{{\thetab}; K}$, where, for $\mathcal{J}_k(K,g_1):= \int_{0}^{1} K(u) K_{g_1}(u) du$,
\begin{equation*}\Gamb_{{\thetab}; K, g_1}:= \frac{\mathcal{J}_k(K,g_1)}{k-1} ({\bf I}_k- \thetab \thetab\pr).
\end{equation*} \Black
\item[(iv)] satisfies, under ${\rm P}_{{\thetab}; g_1}\n$ as $\ny$, the asymptotic linearity property
$$\utDelta_{{\thetab} + n^{-1/2}{\bf t}\n; K}\n- \utDelta_{{\thetab}; K}\n =- \Gamb_{{\thetab}; K, g_1} {\bf t}\n + o_{\rm P}(1)$$
for any bounded sequence ${\bf t}\n$ as described in $(\ref{localt})$.
\end{itemize}
\end{Prop}

The main point of this proposition is part (iv); the preceding three results are necessary for proving that last part. However, they have an interest \emph{per se} as they give the asymptotic properties of the rank-based central sequence $\utDelta_{{\thetab}; K}\n$. Moreover, these results happen to be important when the focus lies on testing procedures for the location parameter $\thetab$. This is part of ongoing research. For the present paper, we are mainly interested in the asymptotic linearity property of $\utDelta_{{\thetab}; K}\n$, more precisely on the asymptotic linearity property of
\begin{equation*}
\utDelta_{\hat{\thetab}; K}\n:=  n^{-1/2} \sum_{i=1}^{n} K \left(\frac{\hat{R}_i}{n+1} \right) \Sb_{\hat{\thetab}}(\Xb_i),
\end{equation*}
where $\hat{R}_i$, $i=1,\ldots,n$, stands for the rank of $\Xb_i'\hat{\thetab}$ among $\Xb_1'\hat{\thetab},\ldots,\Xb_n'\hat{\thetab}$. It is precisely here that Assumption C comes into play: it ensures that the asymptotic linearity property of Proposition~\ref{asymplin}(iv) holds after replacement of ${\bf t}\n$ by the random quantity $n^{1/2}(\hat{\thetab}- \thetab)$ (this can be seen via Lemma 4.4 of \citet{Kre1987}), which eventually entails that
\begin{equation}\label{aslin}
\utDelta_{\hat{\thetab}; K}\n- \utDelta_{{\thetab}; K}\n =- \Gamb_{{\thetab}; K, g_1}  n^{1/2}(\hat{\thetab}- \thetab)+ o_{\rm P}(1)
\end{equation}
as $n\rightarrow\infty$ under ${\rm P}_{{\thetab}; g_1}\n$ for $g_1\in\mathcal{F}$. We draw the reader's attention to the fact that (\ref{localt}) is satisfied when ${\bf t}\n$ is replaced by $n^{1/2}(\hat{\thetab}- \thetab)$.

\subsection{Properties of our $R$-estimators}\label{sec:est2}

Keeping the notation ${\bf A}^{-}$ for the Moore-Penrose inverse of some matrix ${\bf A}$, let
\begin{eqnarray*} 
\tilde{\thetab}_{K; \mathcal{J}_{k}(K,g_1)} &:= &\hat{\thetab}+ n^{-1/2} \Gamb_{\hat\thetab; K, g_1}^{-} \utDelta_{\hat{\thetab}; K}\n \nonumber \\ &=& \hat{\thetab}+ n^{-1/2} \frac{(k-1)}{\mathcal{J}_{k}(K,g_1)} ({\bf I}_{k}- \hat\thetab\hat\thetab\pr)\utDelta_{\hat{\thetab}; K}\n \\ &=& \hat{\thetab}+ n^{-1/2} \frac{(k-1)}{\mathcal{J}_{k}(K,g_1)}  \utDelta_{\hat{\thetab}; K}\n,  \nonumber
\end{eqnarray*}
where the last inequality holds since $\hat\thetab\pr {\bf S}_{\hat \thetab}(\Xb_i)=0$ for all $i=1, \ldots,n$.  The expression of $\tilde{\thetab}_{K; \mathcal{J}_{k}(K,g_1)}$ is quite traditional when one-step estimation is considered. However, using $\tilde{\thetab}_{K; \mathcal{J}_{k}(K,g_1)}$ itself to estimate $\thetab$ is clearly unnatural since $\tilde{\thetab}_{K; \mathcal{J}_{k}(K,g_1)}$ does not belong to $\mathcal{S}^{k-1}$ in general. This is why we propose the one-step $R$-estimator 
\begin{eqnarray*}
\hat{\thetab}_{K; \mathcal{J}_{k}(K,g_1)} &:= \tilde{\thetab}_{K; \mathcal{J}_{k}(K,g_1)}/ \| \tilde{\thetab}_{K; \mathcal{J}_{k}(K,g_1)} \| \in {\cal S}^{k-1}
\end{eqnarray*}
which is a normalized version of  $\tilde{\thetab}_{K; \mathcal{J}_{k}(K,g_1)}$. As it is shown in the sequel, the normalization has no asymptotic cost on the efficiency of the one-step method.  
%

Nevertheless, \Black $\hat{\thetab}_{K;\mathcal{J}_{k}(K,g_1)}$ is not a genuine estimator because it is still a function of the unknown scalar $\mathcal{J}_{k}(K,g_1)$. This cross-information quantity requires to be consistently estimated in order to ensure asymptotic normality of our $R$-estimators. To tackle this problem, we adopt here the idea developed in \citet{HalOjaPai2006} still based on the ULAN property of the model. Define $\tilde{\thetab}(\beta):=\hat{\thetab}+ n^{-1/2} \beta (k-1)  \utDelta_{\hat{\thetab}; K}\n$, $\hat{\thetab}(\beta):= \tilde{\thetab}(\beta)/ \| \tilde{\thetab}(\beta)\|$
and consider the quadratic form
\begin{eqnarray}\label{hfun} \beta \mapsto h\n(\beta)&:=  & (\utDelta_{\hat{\thetab}; K}\n) \pr \Gamb_{\hat\thetab; K}^{-} \utDelta_{\hat{\thetab}(\beta); K}\n \nonumber \\ &=&  \frac{(k-1)}{{\mathcal J}_k(K)}(\utDelta_{\hat{\thetab}; K}\n) \pr \utDelta_{\hat{\thetab}(\beta); K}\n.
\end{eqnarray}
Then, part (iv) of Proposition \ref{asymplin} and the root-$n$ consistency of $\hat{\thetab}(\beta)$ (which follows from the root-$n$ consistency of $\tilde{\thetab}(\beta)$ and the Delta method applied to the mapping $\xb \mapsto \xb/ \| \xb \|$)  imply that, after some direct computations involving~(\ref{aslin}),
$$\utDelta_{\hat{\thetab}(\beta); K}\n- \utDelta_{{\hat{\thetab}}; K}\n =- \Gamb_{{\thetab}; K, g_1}  n^{1/2}(\hat{\thetab}(\beta)- \hat{\thetab})+ o_{\rm P}(1)$$
as $n\rightarrow\infty$ under ${\rm P}_{{\thetab}; g_1}\n$. Moreover, it is clear that $\Gamb_{{\thetab}; K, g_1}  n^{1/2}(\hat{\thetab}(\beta)- \hat{\thetab})=\Gamb_{{\hat{\thetab}}; K, g_1}  n^{1/2}(\hat{\thetab}(\beta)- \hat{\thetab})+o_{\rm P}(1)$ as $n\rightarrow\infty$ under ${\rm P}_{{\thetab}; g_1}\n$. These facts combined with the definition of $\hat{\thetab}(\beta)$ entail that, under ${\rm P}_{\thetab; g_1}\n$ and for $\ny$,
\begin{eqnarray*}
 h\n(\beta) &= &  \frac{(k-1)}{{\mathcal J}_k(K)}  (\utDelta_{\hat{\thetab}; K}\n)\pr  \left( \utDelta_{\hat{\thetab}; K}\n- \Gamb_{\hat\thetab; K, g_1} n^{1/2}(\hat{\thetab}(\beta)-\hat{\thetab}) \right) + o_{\rm P}(1) \nonumber \\
 &=&  \frac{(k-1)}{{\mathcal J}_k(K)}  (\utDelta_{\hat{\thetab}; K}\n)\pr \left( \utDelta_{\hat{\thetab}; K}\n- {\mathcal J}_k(K, g_1)\beta  \utDelta_{\hat{\thetab}; K}\n\right) + o_{\rm P}(1) \nonumber
 \\
 &=&  \frac{(k-1)(1- {\mathcal J}_k(K, g_1)\beta)}{{\mathcal J}_k(K)}  (\utDelta_{\hat{\thetab}; K}\n)\pr \utDelta_{\hat{\thetab}; K}\n  + o_{\rm P}(1) \nonumber
\end{eqnarray*} \Black
where the passage from the first to the second line requires some computations involving again the Delta method, but for the sake of readability we dispense the reader from these calculatory details here; they follow along the same lines as those achieved below. In view of (\ref{hfun}), $ h\n(\beta)$ can be rewritten as
$$\label{hend}h\n(\beta)=(1- {\mathcal J}_k(K, g_1)\beta)h \n(0) + o_{\rm P}(1) $$
under ${\rm P}_{\thetab; g_1}\n$ as $\ny$. Since $h\n(0) >0$, one obtains (the proof is along the same lines as in \citet{HalOjaPai2006}) a consistent estimator of $({\mathcal J}_k(K, g_1))^{-1}$ given by $\hat{\beta}:= {\rm inf} \{ \beta >0 : h\n(\beta) <0 \}$. Therefore, $\hat{\mathcal J}_k(K, g_1):=\hat{\beta}^{-1}$ provides a consistent estimator of the cross-information quantity, and a genuine estimator of $\thetab$ is provided by ${\hat \thetab}_{K; \hat{\mathcal{J}}_{k}(K,g_1)}$. Now, the Delta method and some easy computations show that under  ${\rm P}_{\thetab; g_1}\n$ and for $\ny$
\begin{eqnarray} \label{delta} n^{1/2} ({\hat \thetab}_{K; \hat{\mathcal{J}}_{k}(K,g_1)}- \thetab) &=&
 n^{1/2} \left( \tilde{\thetab}_{K; \hat{\mathcal{J}}_{k}(K,g_1)} / \| \tilde{\thetab}_{K; \hat{\mathcal{J}}_{k}(K,g_1)} \|  - \thetab / \| \thetab \| \right)  \nonumber \\ &=&n^{1/2}  ({\bf I}_k- \thetab \thetab\pr)  \left( \tilde{\thetab}_{K; \hat{\mathcal{J}}_{k}(K,g_1)}   - \thetab  \right) + o_{\rm P}(1),
\end{eqnarray}
where ${\bf I}_k- \thetab \thetab\pr$ is the Jacobian matrix of the mapping ${\xb} \mapsto \xb/ \| \xb \|$ (as already mentioned, the above-used Delta method works similarly) evaluated at $\thetab$. Then, (\ref{delta}), the definition of $ \tilde{\thetab}_{K; \hat{\mathcal{J}}_{k}(K,g_1)}$, part (iv) of Proposition \ref{asymplin}, the consistency of  $\hat{\mathcal{J}}_{k}(K,g_1)$ together with Assumption C entail that, under ${\rm P}_{\thetab; g_1}\n$ and as $\ny$,

\begin{eqnarray} \label{asympeq} n^{1/2} ({\hat \thetab}_{K; \hat{\mathcal{J}}_{k}(K,g_1)}- \thetab) &=& n^{1/2}  ({\bf I}_k- \thetab \thetab\pr)  \left( \tilde{\thetab}_{K; \hat{\mathcal{J}}_{k}(K,g_1)}   - \thetab  \right) +o_{\rm P}(1) \nonumber \\ &=&  n^{1/2}  ({\bf I}_k- \thetab \thetab\pr)  \left(  \hat{\thetab}+ n^{-1/2} \Gamb_{\hat\thetab; K, g_1}^{-} \utDelta_{\hat{\thetab}; K}\n  - \thetab  \right) + o_{\rm P}(1) \nonumber \\ &=&
 ({\bf I}_k- \thetab \thetab\pr) \left( n^{1/2} (\hat{\thetab}- \thetab) + \Gamb_{\thetab; K, g_1}^{-} \utDelta_{{\thetab}; K}\n -  n^{1/2}(\hat\thetab- \thetab) \right) + o_{\rm P}(1) \nonumber \\ &=&
({\bf I}_k- \thetab \thetab\pr) \Gamb_{\thetab; K, g_1}^{-}\utDelta_{{\thetab}; K}\n + o_{\rm P}(1) \nonumber \\ &=& \frac{(k-1)}{{\mathcal J}_k(K,g_1)}\utDelta_{{\thetab}; K}\n + o_{\rm P}(1),
\end{eqnarray}
where the passage from the second to the third line draws upon~(\ref{aslin}). Wrapping up, we obtain the following result which summarizes the asymptotic properties of $\hat{\thetab}_{K;\hat{\mathcal{J}}_{k}(K,g_1)}$.

\begin{Prop} \label{asympnorm} Let Assumptions A, B, C and D hold. Then,

\begin{itemize}

 \item[(i)] under ${\rm P}_{\thetab; g_1}\n$, $n^{1/2} ({\hat \thetab}_{K; \hat{\mathcal{J}}_{k}(K,g_1)}- \thetab) $ is asymptotically normal with mean ${\bf 0}$ and covariance matrix $(k-1) \frac{{\mathcal J}_k(K)}{{\mathcal J}^2_k(K,g_1)}\left({\bf I}_{k}- \thetab \thetab\pr \right)$;
\item[(ii)] for $K(u)=K_{f_1}(u)=\varphi_{f_1}(\tilde{F}_1^{-1}(u)) (1-(\tilde{F}_1^{-1}(u))^2)^{1/2}$, the estimator $\hat{\thetab}_{K; \hat{\mathcal{J}}_{k}(K,f_1)}$ is semi-parametrically efficient under ${\rm P}_{\thetab; f_1}\n$.

\end{itemize}

\end{Prop}

\Black


While part (i) of Proposition \ref{asympnorm} is a direct consequence of (\ref{asympeq}), part (ii) requires more explanations. From (\ref{asympeq}), we obviously have that under ${\rm P}_{\thetab; f_1}\n$

\begin{equation} \label{stepp1} n^{1/2}(\hat{\thetab}_{K_{f_1}; \hat{\mathcal{J}}_{k}(K_{f_1},f_1)}-\thetab)=\frac{(k-1)}{{\mathcal J}_k(K_{f_1},f_1)}\utDelta_{{\thetab}; K_{f_1}}\n + o_{\rm P}(1) \end{equation}
as $\ny$. Now, using the identities ${\mathcal J}_k(K_{f_1},f_1)={\mathcal J}_k(f_1)$ and $\Deltab_{\thetab; K_{f_1}, f_1}\n=\Deltab_{\thetab; f_1}\n$, we obtain via part (i) of Proposition \ref{asymplin} that

\begin{eqnarray} \label{stepp2} \frac{(k-1)}{{\mathcal J}_k(K_{f_1},f_1)}\utDelta_{{\thetab}; K_{f_1}}\n &=& \frac{(k-1)}{{\mathcal J}_k(f_1)}\Deltab_{{\thetab}; {f_1}, }\n + o_{\rm P}(1) \end{eqnarray}
under ${\rm P}_{\thetab; f_1}\n$ as $\ny$. Combining (\ref{stepp1}) and (\ref{stepp2}) (and using again the fact that $\thetab\pr{\Sb}_\thetab({\Xb})=0$), it follows that
\begin{equation*}  n^{1/2}(\hat{\thetab}_{K_{f_1}; \hat{\mathcal{J}}_{k}(K_{f_1},f_1)}-\thetab)= \frac{(k-1)}{{\mathcal J}_k(f_1)}\Deltab_{{\thetab}; {f_1} }\n + o_{\rm P}(1)= \Gamb_{\thetab; f_1}^{-}\Deltab_{{\thetab}; {f_1} }\n + o_{\rm P}(1) \end{equation*}
still under ${\rm P}_{\thetab; f_1}\n$ as $\ny$. This entails that $\hat{\thetab}_{K_{f_1}; \hat{\mathcal{J}}_{k}(K_{f_1},f_1)}$ is asymptotically efficient under ${\rm P}_{\thetab; f_1}\n$, which then finally yields the desired optimality properties.



\Black

Finally note that if the preliminary estimator $\hat{\thetab}$ is rotation-equivariant, meaning that $\hat{\thetab}({\bf O}\Xb_i)= {\bf O}\hat{\thetab}(\Xb_i)$ for any matrix ${\bf O} \in \mathcal{SO}_k$, then $\hat{\thetab}_{K; \hat{\mathcal{J}}_{k}(K,g_1)}({\bf O}\Xb_i)= {\bf O}\hat{\thetab}_{K; \hat{\mathcal{J}}_{k}(K,g_1)}(\Xb_i)$. This implies  that rotation-equivariance of our $R$-estimators is inherited from the preliminary estimator.

\setcounter{equation}{0} 
\section{Asymptotic relative efficiencies and simulation results}\label{sec:ARE}

In this section, we compare the asymptotic and finite-sample performances of the proposed $R$-estimators with those of the spherical mean and spherical median. The asymptotic performances are analyzed in Section~\ref{ARE} on basis of asymptotic relative efficiencies (ARE), while the finite-sample behavior is investigated in Section~\ref{MC} by means of a Monte Carlo study.

\subsection{Asymptotic relative efficiencies}\label{ARE}


As already mentioned in the Introduction,  the traditional  estimators of $\thetab$ in the literature are $M$-estimators. Such an $M$-estimator \Black is defined, for a given function $\rho_0(\xb; \thetab)$,  as the  value $\hat\thetab$ of $\thetab$ which minimizes the objective function
$$\thetab\mapsto\rho(\thetab) := \sum_{i=1}^n \rho_0(\Xb_i; \thetab),$$ \Black
where $\Xb_1,\ldots,\Xb_n$ are spherical observations. 
%
Letting $\rho_0(\xb; \thetab)=:\tilde{\rho}(\xb\pr \thetab)$, special instances of the above are the spherical mean $\hat\thetab_{\rm Mean}$ (maximum likelihood estimator under FVML distributions, obtained by taking $\psi(t):=-\dot{\tilde{\rho}}(t)=2$) and the spherical median  $\hat\thetab_{\rm Median}$ (\citet{Fis1985}, obtained by taking $\psi(t)=(1-t^2)^{-1/2}$).  By Theorem 3.2 in \citet{Cha2004}, an $M$-estimator $\hat{\thetab}_M$ associated with the objective function $\tilde{\rho}(\xb\pr \thetab)$ is such that $n^{1/2}(\hat{\thetab}_M-\thetab)$ is asymptotically normal with mean zero and covariance matrix
$$ (k-1) \frac{{\rm E}[\psi^2(\Xb\pr \thetab) (1-(\Xb\pr \thetab)^2)]}{{\rm E}^2[\psi(\Xb\pr \thetab) \varphi_{g_1}(\Xb\pr \thetab)(1-(\Xb\pr \thetab)^2)]}\left({\bf I}_{k}- \thetab\thetab\pr \right)$$
\Black
under ${\rm P}_{\thetab; g_1}\n$ (expectations are taken under  ${\rm P}_{\thetab; g_1}\n$). Now, let $\hat{\thetab}_1$ and $\hat{\thetab}_2$ be two estimators of the spherical location such that $n^{1/2}(\hat{\thetab}_1-\thetab)$ and $n^{1/2}(\hat{\thetab}_2-\thetab)$ are asymptotically normal with mean zero and covariance matrices $\rho_1({\bf I}-\thetab\thetab\pr)$ and $\rho_2({\bf I}-\thetab\thetab\pr)$, respectively, under ${\rm P}_{\thetab; g_1}\n$. Then, a natural way to compare the asymptotic efficiencies of  $\hat{\thetab}_1$ and $\hat{\thetab}_2$  (still under ${\rm P}_{\thetab; g_1}\n$) is through the ratio ${\rm ARE}_{\thetab; g_1}(\hat\thetab_{1}/\hat{\thetab}_2)=\rho_1/\rho_2$ which we refer to as an ARE. The following result  provides  a general formula for ARE between an $R$-estimator and an $M$-estimator.


\begin{Prop} Let $\hat\thetab_{K; \hat{\mathcal{J}}_{k}(K,g_1)}$ be the $R$-estimator associated with the score function $K$ and let $\hat{\thetab}_M$ be the $M$-estimator associated with the objective function $\rho(\xb\pr \thetab)$. Then
\begin{equation*} \label{AREfor}{\rm ARE}_{\thetab; g_1}(\hat\thetab_{K; \hat{\mathcal{J}}_{k}(K,g_1)}/\hat{\thetab}_M)=\frac{{\rm E}[\psi^2(\Xb\pr \thetab) (1-(\Xb\pr \thetab)^2)]{\mathcal J}^2_k(K,g_1)}{{\rm E}^2[\psi(\Xb\pr \thetab) \varphi_{g_1}(\Xb\pr \thetab)(1-(\Xb\pr \thetab)^2)]{\mathcal J}_k(K)}, \end{equation*}
where ${\rm ARE}_{\thetab; g_1}$ denotes the asymptotic relative efficiency under ${\rm P}_{\thetab; g_1}\n$.
\end{Prop}

In Tables \ref{AREtable1} and  \ref{AREtable2}, we collect numerical values of $\rm ARE_{\thetab; g_1}$ for $k=3$ under various underlying rotationally symmetric densities (namely those described right after Assumption A in Section~\ref{sec:ULAN}). Several one-step $R$-estimators are compared to both the spherical mean and the spherical median: $\hat{\thetab}_{\rm FVML(2)}$ and $\hat{\thetab}_{\rm FVML(6)}$, based on FVML scores (with $\kappa=2$ and $\kappa=6$, respectively), $\hat{\thetab}_{\rm Lin(2)}$ and $\hat{\thetab}_{\rm Lin(4)}$  based on linear scores (associated with linear angular densities with $a=2$ and $a=4$, respectively,  see (\ref{linlog})), $\hat{\thetab}_{\rm Log(2.5)}$ based on logarithmic scores  (associated with a logarithmic angular density with $a=2.5$,  \Black see  (\ref{linlog})) and finally $\hat{\thetab}_{\rm Logis(1,1)}$ and $\hat{\thetab}_{\rm Logis(2,1)}$ based on logistic scores  (associated with logistic angular densities with respectively $a=1, \;  b=1$ and $a=2, \;  b=1$, see (\ref{densities})). \Black

Inspection of Tables  \ref{AREtable1} and  \ref{AREtable2} confirms the theoretical results obtained previously. When based on the score function associated with the underlying density, $R$-estimators are optimal. For example, $\hat{\thetab}_{\rm Lin(2)}$ is the most precise estimator under ${\rm P}_{\thetab; f_{1; {\rm Lin (2)}}}\n$. As expected the spherical mean dominates the $R$-estimators under FVML densities since it is the maximum likelihood estimator in this case. For example, $\hat{\thetab}_{\rm FVML(2)}$ is only just less or equally (under the FVML(2) density) efficient than  the spherical mean under  FVML densities but performs nicely under other densities. In general, the proposed $R$-estimators outperform the spherical median.

\begin{table}[!h]
\begin{center}
{\scriptsize
\begin{tabular}{|c|cccccccc|}
\hline
& \multicolumn{8}{|c|}{AREs with respect to the spherical mean (ARE($\hat{\thetab}_{K}/ \hat{\thetab}_{\rm Mean}$))} \\
\hline
\hline
 Underlying density & $\hat{\thetab}_{\rm FVML(2)}$ &$\hat{\thetab}_{\rm FVML(6)}$& $\hat{\thetab}_{\rm Lin(2)}$ & $\hat{\thetab}_{\rm Lin(4)}$  & $\hat{\thetab}_{\rm Log(2.5)}$& $\hat{\thetab}_{\rm Logis(1,1)}$ & $\hat{\thetab}_{\rm Logis(2,1)}$ & $\hat{\thetab}_{\rm Sq(1.1)}$  \\
 \hline
 \hline
 {\rm FVML(1)} &  0.9744 &0.8787  & 0.9813 & 0.9979 & 0.9027  &0.9321 &  0.7364 & 0.7804 \\
 {\rm FVML(2)} & 1 & 0.9556  &0.9978 & 0.9586 & 0.9749  & 0.9823 &0.8480  & 0.8932\\
 {\rm FVML(6)} & 0.9555 & 1  &0.9381 &0.8517  &  0.9768 & 0.9911 &0.9280  &  0.9771\\
 {\rm Lin(2)} &  1.0539 &  0.9909  &1.0562 & 1.0215 & 1.0212 & 1.0247  &0.8796 & 0.9174 \\
  {\rm Lin(4)} & 0.9709 & 0.8627  & 0.9795 &  1.0128  & 0.8856 &0.9231  &  0.7097  & 0.7083\\
 {\rm Log(2.5)} & 1.1610 & 1.1633 & 1.1514  &1.0413  & 1.1908 & 1.1625  &  1.0951  & 1.1376 \\
  {\rm Log(4)} & 1.0182 & 0.9216 & 1.0261& 1.0347 & 0.9503 & 0.9741 &  0.7851  & 0.8226\\
   {\rm Logis(1,1)} & 1.0768  & 1.0865 & 1.0635 & 0.9991 & 1.0701  & 1.0962 & 0.9778 &  1.0349\\
    {\rm Logis(2,1)} &  1.3182 & 1.4426 & 1.2946 & 1.0893 & 1.4294 &1.3865 & 1.5544  & 1.4680\\
    {\rm Sq(1.1)} &  1.2303  &  1.3460 &  1.1964 & 1.0264  &  1.3158 & 1.3004 & 1.3009  & 1.3774 \\
    \hline

\end{tabular}
}
\end{center}
\caption{\small Asymptotic relative efficiencies of $R$-estimators with respect to the spherical mean under various $3$-dimensional rotationally symmetric densities.
}
\label{AREtable1}
\end{table}

\begin{table}[!h]
\begin{center}
{\scriptsize
\begin{tabular}{|c|cccccccc|}
\hline
& \multicolumn{8}{|c|}{AREs with respect to the spherical median (ARE($\hat{\thetab}_{K}/ \hat{\thetab}_{\rm Median}$))} \\
\hline
\hline
 Underlying density & $\hat{\thetab}_{\rm FVML(2)}$& $\hat{\thetab}_{\rm FVML(6)}$ & $\hat{\thetab}_{\rm Lin(2)}$ & $\hat{\thetab}_{\rm Lin(4)}$  & $\hat{\thetab}_{\rm Log(2.5)}$& $\hat{\thetab}_{\rm Logis(1,1)}$ & $\hat{\thetab}_{\rm Logis(2,1)}$ & $\hat{\thetab}_{\rm Sq(1.1)}$  \\
 \hline
 \hline
 {\rm FVML(1)} &  1.0691 & 0.9641 &1.0766 & 1.0949 & 0.9904  &1.0226 &  0.8079 & 0.8562 \\
 {\rm FVML(2)} &1.1321 & 1.0819 & 1.1297 & 1.0853 & 1.1038  & 1.1121 &  0.9601  &  1.0112 \\
 {\rm FVML(6)} & 1.1632 & 1.2174 & 1.1421 & 1.0369  &  1.1891 & 1.2065&1.1297 &  1.1895 \\
 {\rm Lin(2)} &  1.1391 & 1.0709  & 1.1415 &  1.1041 & 1.1037  &1.1075  &0.9507 & 0.9915  \\
  {\rm Lin(4)} &   1.0493 & 0.9324  &  1.0587 &  1.0946  & 0.9571 &0.9977  &   0.7671 & 0.8157 \\
 {\rm Log(2.5)} & 1.2171 & 1.2195 & 1.2071  & 1.0917  & 1.2484& 1.2188  & 1.1481 & 1.1926\\
  {\rm Log(4)} & 1.0900 & 0.9865 & 1.0984& 1.1076 & 1.0173 & 1.0427 &  0.8403 & 1.0228 \\
   {\rm Logis(1,1)} & 1.1264   &  1.1365 & 1.1125 & 1.0451 & 1.1194  &  1.1467 & 1.0228 & 1.0826  \\
    {\rm Logis(2,1)} &   1.4497&  1.5868 & 1.4238 &1.1980 &1.5721 &1.5249 &  1.7095 &  1.6145 \\
     {\rm Sq(1.1)} & 1.2282   & 1.3436   & 1.1943  & 1.0246  &   1.3135 & 1.2981  & 1.2986 & 1.3750 \\
        \hline

\end{tabular}
}
\end{center}
\caption{\small Asymptotic relative efficiencies of $R$-estimators with respect to the spherical median under various $3$-dimensional rotationally symmetric densities.
}
\label{AREtable2}
\end{table}

\subsection{Monte Carlo study}\label{MC}

We now discuss the finite-sample behavior of different $R$-estimators of the spherical location. For this purpose, we have generated $M=1000$ samples  from various 3-dimensional  ($k=3$) rotationally symmetric distributions: (i) the FVML(2) and FVML(4) distributions,  (ii) the linear distribution with $a=2$ and $a=4$ (see (\ref{linlog})) and (iii) the \emph{square root} distribution ${\rm Sq}(1.1)$ associated with an angular density of the form
$$ f_1(t):= \sqrt{t+a}, \quad {\rm with} \; a=1.1.$$ The true location parameter is $\thetab=(\sqrt{2}/2, \sqrt{2}/2,0)\pr$.
For each replication, the spherical median $\hat{\thetab}_{\rm Median}$,  the spherical mean $\hat{\thetab}_{\rm Mean}$, the FVML-score based $R$-estimators $\hat{\thetab}_{\rm FVML(2)}$ and $\hat{\thetab}_{\rm FVML(4)}$, the linear-score based $R$-estimators $\hat{\thetab}_{\rm Lin(2)}$ and $\hat{\thetab}_{\rm Lin(4)}$  and finally the ${\rm Sq}(1.1)$-score based $R$-estimator $\hat{\thetab}_{{\rm Sq}(1.1)}$ have been computed. The  preliminary estimators used in the construction of all our one-step $R$-estimators are $\hat{\thetab}_{\rm Median}$ and $\hat{\thetab}_{\rm Mean}$, respectively.

In Table \ref{MSE1}, we report  the Euclidean norm of ${\bf m}=({m}_1, {m}_2, {m}_3)\pr $ where, letting $\hat{\theta}_i^{(j)}$ stand for the $i$th component of an estimator computed from the $j$th replication and $\theta_i$ for the $i$th component of $\thetab$,
$$m_i:=\frac{1}{M} \sum_{j=1}^M (\hat{\theta}_i^{(j)}- \theta_i)^2, \;  i=1,2,3,$$
is computed still for each of the aforementioned estimators and sample sizes $n=100$, $n=500$ and $n=1000$. Simulation results mostly confirm the ARE rankings. 
Under all the distributions considered, the optimality of the $R$-estimators based on correctly
specified densities is verified. In the FVML case, the spherical mean is efficient as expected, but the $R$-estimators based on FVML scores are reasonable competitors. In general, the efficiency of our $R$-estimators becomes better with respect to the spherical mean under departures from the FVML case, especially under the ${\rm Sq}(1.1)$ density. The spherical median is clearly dominated by the other estimators. The results also show that, as $n$ increases, the influence of the choice of the preliminary estimator wanes; this confirms the fact that the asymptotic behavior of the $R$-estimators does not depend on this choice (see Assumption C). \vspace{4mm}

\Black

\begin{table}[htbp]
\begin{center}
{\tiny
\begin{tabular}{|cc|cc|cc|cc|}
\hline
 & & & & & & & \\
 \multicolumn{2}{|c|}{\tiny{\rm Sample size}}&  \multicolumn{2}{c|}{{\tiny $n=100$}} & \multicolumn{2}{c|}{{\tiny $n=500$}} & \multicolumn{2}{c|}{{\tiny $n=1000$}}   \\[2mm]
\hline
& & & & & & & \\
 \multicolumn{2}{|c|}{\tiny{\rm Preliminary estimator}}&$\hat{\thetab}_{\rm Mean}$ & $\hat{\thetab}_{\rm Median}$ & $\hat{\thetab}_{\rm Mean}$ & $\hat{\thetab}_{\rm Median}$ & $\hat{\thetab}_{\rm Mean}$ & $\hat{\thetab}_{\rm Median}$  \\
& & & & & & & \\
{\tiny{\rm Actual density}}&{\tiny{\rm Estimators}} & & & & & & \\[1mm]
\hline
& \hspace{-2mm}         ${\hat \thetab}_{{\rm FVML} (2)}$ & .12846 & .12813 & .00528 & .00528 & .00135 & .00134 \\
&  \hspace{-2mm}  ${\hat \thetab}_{{\rm FVML} (4)}$ & .13724 & .13779 &  .00550 &  .00551 &.00143 &.00142\\
& \hspace{-5.5mm} ${\hat \thetab}_{{\rm Lin} (2)}$ & .12818 & .12836& .00531 & .00532 & .00135 & .00134 \\
{\rm FVML} (2)&  \hspace{-4.5mm}${\hat \thetab}_{{\rm Lin} (4)}$ &.13540 & .13568 & .00576 & .00577 & .00147 & .00146 \\
&  \hspace{-4mm}  ${\hat \thetab}_{{\rm Sq} (1.1)}$ & .16916 & .17003 & .00647 & .00648 & .00170 & .00170\\
&  \hspace{-6mm} ${\hat \thetab}_{{\rm Mean}}$ & .12686 & ------ & .00528 &------  & .00135 & ------   \\
&  \hspace{-4mm}  ${\hat \thetab}_{{\rm Median}}$ & ------  & .16649 & ------  & .00687 & ------  & .00172 \\[1mm]
\hline
& \hspace{-2mm}          ${\hat \thetab}_{{\rm FVML} (2)}$ & .01815 & 	.01810 & .00072 & .00072 & .00019 & .00018 \\
&\hspace{-2mm}      ${\hat \thetab}_{{\rm FVML} (4)}$ & .01756 & .01757 &  .00069 &  .00069 &.00018 &.00017\\
& \hspace{-5.5mm}   ${\hat \thetab}_{{\rm Lin} (2)}$ & .01852 & .01849& .00075 & .00074 & .00019 & .00019 \\
{\rm FVML} (4)& \hspace{-5.5mm}   ${\hat \thetab}_{{\rm Lin} (4)}$ &.02158 & .02160 & .00087 & .00087 & .00022 & .00022 \\
&\hspace{-4mm}   ${\hat \thetab}_{{\rm Sq} (1.1)}$ & .01898 & .01909 & .00074 & .00074 & .00019 & .00018\\
& \hspace{-6mm}  ${\hat \thetab}_{{\rm Mean}}$ & .01742 & ------ & .00069 &------  & .00018 & ------   \\
& \hspace{-4mm}  ${\hat \thetab}_{{\rm Median}}$ & ------  & .02481 & ------  & .00098 & ------  & .00023 \\[1mm]
\hline
&\hspace{-2mm}           ${\hat \thetab}_{{\rm FVML} (2)}$ & 16.40147 & 15.88246 & .80516 & .80281 & .20099 & .20068 \\
& \hspace{-2mm}     ${\hat \thetab}_{{\rm FVML} (4)}$ & 17.61341 & 17.41290 &  .85555 &  .85874 &.21485 &.21566\\
& \hspace{-5.5mm}  ${\hat \thetab}_{{\rm Lin} (2)}$ & 16.27424 & 16.73361& .80201 & .80703 & .20081 & .20137 \\
{\rm Lin} (2)& \hspace{-5.5mm}  ${\hat \thetab}_{{\rm Lin} (4)}$ &16.13449 & 16.64150 & .85463 & .85917 & .21771 & .21827 \\
& \hspace{-4mm}  ${\hat \thetab}_{{\rm Sq} (1.1)}$ & 20.41046 & 20.14262 & 1.05501 & 1.06998 & .25629 & .25712\\
&  \hspace{-6mm}  ${\hat \thetab}_{{\rm Mean}}$ & 16.62219 & ------ & .89528 &------  & .22810 & ------   \\
&\hspace{-4mm} ${\hat \thetab}_{{\rm Median}}$ & ------  & 19.16783  & ------  & 1.06145 & ------  & .26050 \\[1mm]
\hline
& \hspace{-2mm}         ${\hat \thetab}_{{\rm FVML} (2)}$ & 201.7671 & 186.5235 & 16.30444 & 15.85266 & 3.75891 & 3.73754 \\
&\hspace{-2mm}    ${\hat \thetab}_{{\rm FVML} (4)}$ & 212.7127 & 212.5497 &  18.71233 &  18.35010 & 4.35718 & 4.46196 \\
& \hspace{-5.5mm} ${\hat \thetab}_{{\rm Lin} (2)}$ & 199.7272 & 203.3786 & 15.94977 & 16.21466 & 3.68828 & 3.71774 \\
{\rm Lin} (4) & \hspace{-5.5mm}  ${\hat \thetab}_{{\rm Lin} (4)}$ & 187.5686 & 195.9556& 14.23143 & 14.57779 & 3.38679 & 3.42702 \\
&\hspace{-4mm}  ${\hat \thetab}_{{\rm Sq} (1.1)}$ & 231.3489 & 229.2380 & 22.61492 & .23.40273 & 5.64923 & 5.92582 \\
&  \hspace{-6mm}  ${\hat \thetab}_{{\rm Mean}}$ & 186.9112 & ------ & 14.26441 &------  & 3.43922 & ------   \\
& \hspace{-4mm} ${\hat \thetab}_{{\rm Median}}$ & ------  & 195.7693 & ------  & 17.05978 & ------  & 3.94507 \\[1mm]
\hline
& \hspace{-2mm}           ${\hat \thetab}_{{\rm FVML} (2)}$ & 13.28350 & 13.13711 &  .50958& .50620 & .12659 & .12613  \\
&\hspace{-2mm}     ${\hat \thetab}_{{\rm FVML} (4)}$ & 13.05349 & 13.54705 & .44770 & .44563 & .10990 & .11025 \\
&\hspace{-5.5mm} ${\hat \thetab}_{{\rm Lin} (2)}$ & 13.51480 & 13.93615 & .53481 & .53583 & .13304 & .13324  \\
{\rm Sq} (1.1) & \hspace{-5.5mm}  ${\hat \thetab}_{{\rm Lin} (4)}$& 15.82969 &  16.31837& .71788 & .72184 &.18107 & .18156  \\
&\hspace{-4mm}   ${\hat \thetab}_{{\rm Sq} (1.1)}$ & 13.57166 & 14.35157 & .42719 &  .42348 & .10430 & .10487 \\
& \hspace{-6mm}   ${\hat \thetab}_{{\rm Mean}}$ & 16.92412 &------& .77354  & ------ &  .19337 & ------   \\
&\hspace{-4mm} ${\hat \thetab}_{{\rm Median}}$ &------& 18.55458 & ------ &.78308 & ------  & .19223  \\[1mm]
\hline
\end{tabular}
\Black}
\end{center}
\caption{\small MSE of various $R$-estimators and of the spherical mean and spherical median. For the sake of clarity, the results have been multiplied by $10^3$. $R$-estimators are computed by taking both the spherical median and the spherical mean as preliminary estimators. \Black}
\label{MSE1}
\end{table}

\Black

\setcounter{equation}{0} 
\section{A real-data application}\label{empap}

We now apply our $R$-estimators on a real-data example. The data consists of 26 measurements of magnetic remanence made on samples collected from Palaeozoic red-beds in Argentina, and has already been used in \citet{Emb70} and \citet{FisLewEmb87}.
 The purpose of the study  is to determine the origin of natural remanent magnetization in red-beds. While  it is  reasonable to assume that the underlying distribution associated with this data is unimodal (because a single component of magnetization is present, see \citet{FisLewEmb87}), a complete specification of the underlying distribution can not be justified, and hence semi-parametric methods are required.

\begin{figure}[htbp!]
\begin{center}
\includegraphics[height=8cm, width=8cm]{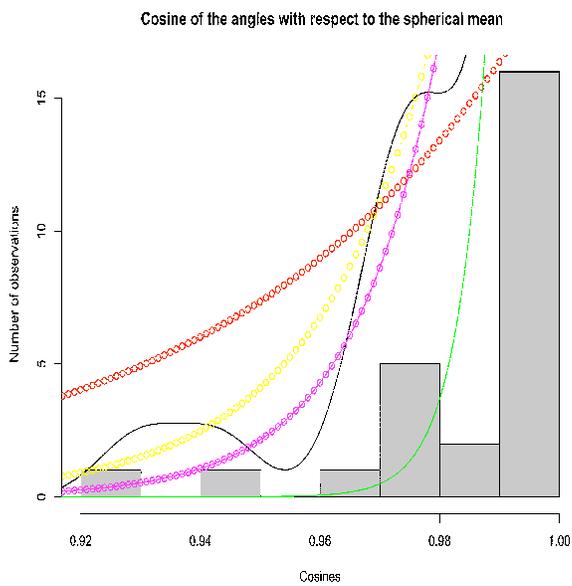}
\end{center}
\vspace{-0mm}
\caption{\small Histogram of the cosines $\Xb_1\pr{\hat \thetab}_{\rm Mean},\ldots, \Xb_{26}\pr{\hat \thetab}_{\rm Mean}$ of the Palaeozoic data. The red dotted line,  the yellow dotted line, the magenta ``overplotted" line and the green line are plots of the FVML densities with $\kappa=20$, $\kappa=50$, $\kappa=69.544$ and $\kappa=200$ respectively. The black line represents a Gaussian kernel density estimator.}
\label{cosines}
\end{figure}

First, we computed the spherical mean ${\hat \thetab}_{\rm Mean}=(.3187373, .4924234, -.8098924)\pr$ and the spherical median ${\hat \thetab}_{\rm Median}=(.3129147,  .486036, -.8159984)\pr$. On basis of this, we provide a histogram of the cosines $\Xb_1\pr{\hat \thetab}_{\rm mean}, \ldots, \Xb_{26}\pr{\hat \thetab}_{\rm mean}$; see Figure \ref{cosines} (the histogram obtained from the cosines $\Xb_1\pr{\hat \thetab}_{\rm Median}, \ldots, \Xb_{26}\pr{\hat \thetab}_{\rm Median}$ has a very similar shape). A visual inspection of the histogram pleads in favor of a FVML score-based $R$-estimation (the black line represents a Gaussian kernel density estimator). Now, even if the FVML family is the target family of densities, the concentration parameter has to be chosen to perform our one-step $R$-estimation. The FVML maximum likelihood estimator of $\kappa$ is given by $\hat{\kappa}_{\rm MLE}=69.544$ (see, e.g., \citet{Ko1992}). 

\begin{figure}[htbp!]
\begin{center}
\includegraphics[height=7cm, width=7cm]{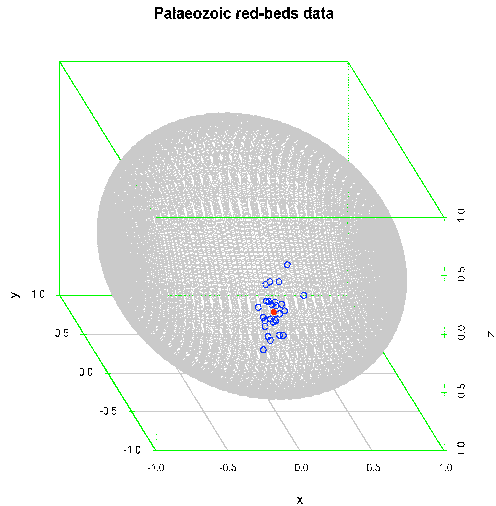}
\end{center}
\vspace{-0mm}
\caption{\small The Palaeozoic data (blue circles) and ${\hat \thetab}_{\kappa=50}$ (red point). The data consists of 26 measurements of magnetic remanence.}
\label{data}
\end{figure}

We computed FVML-score based $R$-estimators with $\kappa=20$, \Black $\kappa=50$, $\kappa=200$ and $\kappa=69.544$ (we took ${\hat \thetab}_{\rm Median}$ as preliminary estimator). We obtained ${\hat \thetab}_{\kappa=20}=(.3152545, .4947636, -.8099027)\pr$, ${\hat \thetab}_{\kappa=50}=(.3156331, .4949186, -.8096652)\pr$, ${\hat \thetab}_{\kappa=200}=(.3156907, .4948861, -.8096624)\pr$ and ${\hat \thetab}_{\kappa=69.544}=(0.3156517, 0.4949093, -0.8096636)\pr$. The data is illustrated in Figure \ref{data}. Data points are represented by blue circles. The red point is ${\hat \thetab}_{\kappa=50}$.

Of course, one could argue that if the FVML specification is chosen, the spherical mean ${\hat \thetab}_{\rm Mean}$ is the efficient estimator and, as a consequence, has to be used for the estimation of $\thetab$. We insist here on the fact that we do not  specify a FMVL distribution but we rather choose the FVML family as a target. As shown in the ARE results (see Table~\ref{AREtable1}), FVML-score based estimators are \emph{robust-efficient} in the sense that if the underlying distribution is not a FVML one, they can be more efficient than the spherical mean.

The empirical example provided in this section raises two  important open questions which are  beyond the scope of the present paper. First,  is $\hat{\thetab}_{\hat{\kappa}}$ (the $R$-estimator based on a FVML score with an estimated concentration parameter) as efficient as $\hat{\thetab}_{\rm Mean}$ under any FVML distribution, irrespective of the underlying concentration? Secondly, is it more efficient than the same spherical mean outside of the FVML case? The answer to these questions could be obtained by considering a location/scale  model in which we would quantify the substitution of the scale parameter by a root-$n$ consistent estimate. This problem is currently under investigation.

\appendix
\section{Proofs}

\noindent {\bf Proof of Proposition \ref{ULAN}.}
Since the LAN property is stated with respect to $\etab$, the density function (\ref{density}) is denoted by $ f_{\etab}$ in this proof but remains the same function. Our proof relies on Lemma 1 of \citet{Swensen1985}--more precisely, on its extension in \citet{GaHa95}. The sufficient conditions for LAN in those results readily follow from standard arguments (hence are left to the reader), once it is shown that $\etab \mapsto f_{\etab}^{1/2}(\xb)$ is differentiable in quadratic mean.

In what follows, all $o(\| \cdot \|)$ or $O(\| \cdot \|)$ quantities are taken as $\| \cdot \| \rightarrow 0$. Denoting by
$${\rm grad}_{\etab} f_{\etab}^{1/2}(\xb):= \frac{1}{2}  f_{\etab}^{1/2}(\xb) \varphi_{f_1}(\xb\pr \hbar(\etab)) {D}\hbar{(\etab)}\pr \xb$$ the gradient of the square root of the density $f_{\etab}(\xb)$, quadratic mean differentiability holds if
\begin{equation} \label{qmd} \int_{\mathcal{S}^{k-1} } \left\{ f_{\etab+ {\bf e}}^{1/2}(\xb)-f_{\etab}^{1/2}(\xb)-{\bf e}\pr {\rm grad}_{\etab} f_{\etab}^{1/2}(\xb)\right\}^2  d\sigma(\xb) \end{equation} is $o(\| {\bf e} \|^2)$ for ${\bf e}\in\R^{k-1}$. Obviously, since $\etab \mapsto \hbar(\etab)$ is differentiable, we have that $\xb'\hbar(\etab + {\bf e})- \xb'\hbar(\etab)=\xb'{D}\hbar{(\etab)} {\bf e} + o(\| {\bf e} \|)$ for all $\xb\in\mathcal{S}^{k-1}$. This implies that the integral (\ref{qmd}) takes the form
\begin{eqnarray} \label{qmdbis} \nonumber c_{k; f_1} \int_{\mathcal{S}^{k-1} } \Big\{ f_1^{1/2}(\xb\pr \hbar(\etab) + \xb\pr {D}\hbar{(\etab)} {\bf e} + o(\| {\bf e} \|) )-  f_1^{1/2}(\xb\pr \hbar(\etab)) \\   -\frac{1}{2} f_{1}^{1/2}(\xb\pr \hbar(\etab)) \varphi_{f_1}(\xb\pr \hbar(\etab)) \xb\pr{D}\hbar{(\etab)} {\bf e} \Big\}^2  d\sigma(\xb). \end{eqnarray}
Now, since $f^{1/2}$ inherits absolute continuity from $f$, $f^{1/2}$ is differentiable almost everywhere on $[-1,1]$. Consequently, for almost all $\xb$ and any perturbation $s\in\R$,
$$ f_1^{1/2}(\xb\pr \hbar(\etab) + s) -  f_1^{1/2}(\xb\pr \hbar(\etab)) = \frac{1}{2} f_{1}^{1/2}(\xb\pr \hbar(\etab)) \varphi_{f_1}(\xb\pr \hbar(\etab)) s + o(s),$$
so that, using the fact that $\sup_{\xb \in \mathcal{S}^{k-1}} \vert\xb\pr{D}\hbar{(\etab)}  {\bf e} \vert \leq C \| {\bf e}\|$ for some positive constant $C$, we have that
$$ \sup_{\xb \in \mathcal{S}^{k-1} } \left|f_1^{1/2}(\xb\pr \hbar(\etab) + \xb\pr{D}\hbar(\etab)  {\bf e} + o(\| {\bf e} \|) )-  f_1^{1/2}(\xb\pr \hbar(\etab)) - \frac{1}{2} f_{1}^{1/2}(\xb\pr \hbar(\etab)) \varphi_{f_1}(\xb\pr \hbar(\etab)) \xb\pr{D}\hbar{(\etab)} {\bf e} \right|$$
is $o(\|{\bf e}\|)$ uniformly in $\xb$. Consequently, the integrand in (\ref{qmdbis}) is $o(\| {\bf e}\|^2)$ uniformly in $\xb$. The result follows since $\int_{\mathcal{S}^{k-1} } d\sigma(\xb)=2 \pi^{k/2}/\Gamma(\frac{k}{2}) <\infty$.

\cqfd \vspace{4mm}

\noindent {\bf Proof of Proposition \ref{ULANbis}.} In this proof, we only provide the expressions of $\Deltab_{\thetab; f_1}\n$ and $\Gamb_{\thetab; f_1}$ since ULAN directly follows from the combination of  Lemma \ref{LElemme} and Proposition \ref{ULAN}.

From Lemma \ref{LElemme}, we obtain that the family  $\left\{ {\rm P}\n_{{\thetab}; f_1} \;  \vert \; {\thetab} \in {\mathcal S}^{k-1} \right\}$ is also ULAN with central sequence $$\Deltab\n_{{\thetab}; f_1}:=n^{-1/2} D\lbar(\etab) ((D\lbar (\etab ) )\pr D\lbar (\etab ))^{-1}(D\lbar (\etab ))\pr  \sum_{i=1}^{n} \varphi_{f_1}(\Xb_i \pr \thetab) (1-(\Xb_i \pr \thetab)^2)^{1/2}  \Sb_\thetab(\Xb_i) $$ and Fisher information matrix $$ \Gamb_{\thetab; f_1}=\frac{\mathcal{J}_k({f_1})}{k-1}  D\lbar(\etab) ((D\lbar (\etab ) )\pr D\lbar (\etab ))^{-1}(D\lbar (\etab ))\pr.$$

Next, note that if $\thetab=\lbar(\etab)$ for some $\etab \in \R^{k-1}$ such that $D\lbar$ is full-rank in the vicinity of $\etab$, then, clearly,  $\thetab=\lbar(\etab)={\bf O}\hbar(\etab)$, with $\hbar$ defined as in (\ref{param1}), for some rotation ${\bf O} \in \mathcal{SO}_k$. After some easy computations, we obtain that (putting $\tilde{\thetab}={\bf O}\pr \thetab=\hbar(\etab)$)
\begin{eqnarray*} D\lbar(\etab) ((D\lbar (\etab ) )\pr D\lbar (\etab ))^{-1}(D\lbar (\etab ))\pr &=& {\bf O} D\hbar(\etab) ((D\hbar (\etab ) )\pr D\hbar (\etab ))^{-1}(D\hbar (\etab ))\pr {\bf O}\pr \\ &=& {\bf O} ({\bf I}_k- {\tilde\thetab} {\tilde\thetab}\pr) {\bf O}\pr \\ &=& {\bf I}_k- {\thetab} {\thetab}\pr.
\end{eqnarray*}

Wrapping up, we see that, combining  Proposition \ref{ULAN} and Lemma \ref{LElemme},  the family $\left\{ {\rm P}\n_{{\thetab}; f_1} \;  \vert \; {\thetab} \in {\mathcal S}^{k-1} \right\}$ (in the $\thetab$-parameterization) is ULAN with central sequence
 \begin{eqnarray*} \Deltab\n_{{\thetab}; f_1} &= &n^{-1/2}({\bf I}_k- \thetab \thetab\pr)  \sum_{i=1}^{n} \varphi_{f_1}(\Xb_i \pr \thetab) (1-(\Xb_i \pr \thetab)^2)^{1/2}  \Sb_\thetab(\Xb_i)\\ &=& n^{-1/2} \sum_{i=1}^{n} \varphi_{f_1}(\Xb_i \pr \thetab) (1-(\Xb_i \pr \thetab)^2)^{1/2}  \Sb_\thetab(\Xb_i) \end{eqnarray*} and Fisher information matrix
\begin{equation*}  \Gamb_{\thetab; f_1}=\frac{\mathcal{J}_k({f_1})}{k-1}({\bf I}_k- \thetab \thetab\pr). \end{equation*}
\cqfd \vspace{4mm}

\noindent {\bf Proof of Proposition \ref{asymplin}.}
Part (i) follows easily from H\'{a}jek's classical result for linear signed-rank statistics (see \citet{HaSi}). Parts (ii) and (iii) are consequences of part (i), the multivariate central limit theorem and Le Cam's third lemma. We therefore only prove in detail part (iv) of the Proposition.

 For the sake of simplicity, we let $\thetab\n=\thetab+n^{-1/2}{\bf t}$ be the perturbed spherical location. In the sequel, we put
$\Ub_i\n:=\Xb_i-(\Xb_i\pr \thetab\n) \thetab\n$, $\Ub_i^{0}:=\Xb_i-(\Xb_i\pr \thetab) \thetab$, $\Sb_{\thetab\n}(\Xb_i):=\Ub_i\n / \vert \vert \Ub_i\n \vert \vert $ and $\Sb_{\thetab}(\Xb_i):=\Ub_{i}^{0} / \vert \vert \Ub_i^{0}\vert \vert $. We then have the  following lemma.

\begin{Lem}\label{signconver} For all $i\in\{1,\ldots,n\}$, we have that  $\vert \vert \Sb_{\thetab\n}(\Xb_i)-\Sb_{\thetab}(\Xb_i)\vert \vert$ is $o_{\rm P}(1)$ under ${\rm P}_{\thetab; f_1}\n$ with $f_1 \in \mathcal{F}$ as $\ny$.
\end{Lem}

\noindent {\bf Proof of Lemma \ref{signconver}}. First note that \begin{eqnarray} \label{triin} \vert \vert \Ub_i\n -  \Ub_i^{0} \vert \vert &=& \| (\Xb_i- (\Xb_i\pr \thetab\n)\thetab\n)- (\Xb_i- (\Xb_i\pr \thetab)\thetab) \| \nonumber \\ &=& \| (\Xb_i\pr \thetab) \thetab - (\Xb_i\pr \thetab) \thetab\n +(\Xb_i\pr \thetab) \thetab\n - (\Xb_i\pr \thetab \n) \thetab\n \| \nonumber \\ &\leq & \| (\Xb_i\pr \thetab)(\thetab- \thetab\n)\| + \| \Xb_i \pr (\thetab- \thetab\n) \thetab\n \|.
\end{eqnarray}
Both terms in (\ref{triin}) are clearly $o_{\rm P}(1)$ as $\ny$.  Consequently, we have that \linebreak
$ \| \Sb_{\thetab\n} (\Xb_i)- \Sb_{\thetab}(\Xb_i) \|  \leq 2 \;  \| \Ub_i^0\|^{-1}  (\| \Ub_i\n - \Ub_i^0\|)$
is $o_{\rm P}(1)$ under ${\rm P}_{\thetab; f_1}\n$ as $\ny$. 

\,\hfill \cqfd \vspace{4mm}

\noindent {\bf Proof of part (iv).} From part (i) we know that $\utDelta_{{\thetab}; K}\n-\Deltab_{{\thetab}; K, g_1}\n=o_{\rm P}(1)$ under ${\rm P}_{\thetab; g_1}\n$ as $\ny$. Similarly, $\utDelta_{{\thetab}\n; K}\n-\Deltab_{{\thetab}\n; K, g_1}\n=o_{\rm P}(1)$ under ${\rm P}_{\thetab\n; g_1}\n$  as $\ny$. Hence, from contiguity, $\utDelta_{{\thetab}\n; K}\n-\Deltab_{{\thetab}\n; K, g_1}\n$ is also $o_{\rm P}(1)$ under  ${\rm P}_{\thetab; g_1}\n$ as $\ny$. This entails that the claim holds if
$$\Deltab_{{\thetab}\n; K, g_1}\n-\Deltab_{{\thetab}; K, g_1}\n+\Gamb_{\thetab; K, g_1} {\bf t}\n$$
is $o_{\rm P}(1)$  under ${\rm P}_{\thetab; g_1}\n$ as $\ny$. Consequently, the result follows if we can show that
\begin{itemize}
\item[{(a)}] $\Deltab_{{\thetab}\n; K, g_1}\n- \Deltab_{{\thetab}; K, g_1}\n-{\rm E}[ \Deltab_{{\thetab}\n; K, g_1}\n]$
is $o_{L^2}(1)$ under ${\rm P}_{\thetab;g_1}\n$ as $\ny$, and that
\item[{(b)}] ${\rm E} \left[ \Deltab_{{\thetab}\n; K, g_1}\n\right] + \Gamb_{\thetab; K,g_1}{\bf t}\n$
is $o(1)$ under ${\rm P}_{\thetab;g_1}\n$ as $\ny$.
\end{itemize}
We first prove \noindent  {(a)}. Using the fact that $\Xb_i\pr \thetab$ and $ \Sb_{\thetab}(\Xb_i)$ are independent (as mentioned in the Introduction), we have (the expectation is taken under ${\rm P}\n_{\thetab; g_1}$)
$${\rm E} \left[ \Deltab_{{\thetab}; K, g_1}\n\right]= n^{-1/2} \sum_{i=1}^n {\rm E}\left[K(\tilde{G}_1(\Xb_i\pr \thetab)) \right] {\rm E}\left[ \Sb_{\thetab}(\Xb_i)\right]= {\bf 0},$$
since,  under ${\rm P}_{\thetab;g_1}\n$,  the sign $ \Sb_{\thetab}(\Xb_i)$ is uniformly distributed on $\mathcal{S}^{k-1}(\thetab^{\perp})$. Now, let ${\bf D}\n:=n^{-1/2} \sum_{i=1}^n \left({\bf T}_i\n -{\rm E}[{\bf T}_i\n]\right)$, where ${\bf T}_i\n:=K(\tilde{G}_1(\Xb_i\pr \thetab\n)) \Sb_{\thetab\n}(\Xb_i)-K(\tilde{G}_1(\Xb_i\pr \thetab)) \Sb_{\thetab}(\Xb_i).$ Clearly, ${\bf D}\n$ corresponds exactly to the expression in (a), since ${\rm E} [ \Deltab_{{\thetab}; K, g_1}\n]=0$ as explained above. Using i.i.d.-ness and standard algebra arguments, we have that, under ${\rm P}_{\thetab;g_1}\n$,
\begin{eqnarray*} {\rm E}[ \| {\bf D}\n \|^2] & = & n^{-1} {\rm E} \left[ \left\| \sum_{i=1}^n {\bf T}_i\n -{\rm E}[{\bf T}_i\n]  \right\|^2 \right]  \\ &=& n^{-1} {\rm tr} \left[{\rm Var} \left[ \sum_{i=1}^n {\bf T}_i\n -{\rm E}[{\bf T}_i\n]  \right] \right] \\ &= &  {\rm E} [\| {\bf T}_1\n \|^2].
\end{eqnarray*}
Therefore, it remains to show that $ {\rm E} [\| {\bf T}_1\n \|^2]$ is $o(1)$ as $\ny$. We have that
\begin{eqnarray*}
\| {\bf T}_1\n \|^2   &=& \Big\| \left(K(\tilde{G}_1(\Xb_1\pr \thetab\n))-K(\tilde{G}_1(\Xb_1\pr \thetab))\right)  \Sb_{\thetab\n}(\Xb_1) \\ & & \hspace{30mm}+K(\tilde{G}_1(\Xb_1\pr \thetab)) \left(\Sb_{\thetab\n}(\Xb_1)- \Sb_{\thetab}(\Xb_1) \right)\Big\|^2 \\ & \leq &  \; 2 \left| K(\tilde{G}_1(\Xb_1\pr \thetab\n))-K(\tilde{G}_1(\Xb_1\pr \thetab))\right|^2 \|\Sb_{\thetab\n}(\Xb_1) \|^2 \\ & & \hspace{30mm} + 2 \;  K^2(\tilde{G}_1(\Xb_1\pr \thetab)) \| \Sb_{\thetab\n}(\Xb_1)- \Sb_{\thetab}(\Xb_1) \|^2.
\end{eqnarray*}
The continuity of $K \circ  \tilde{G}_1$ together with the fact that $\Xb_1\pr \thetab - \Xb_1\pr \thetab\n$ is $o_{\rm P}(1)$ (under ${\rm P}_{\thetab;g_1}\n$ as $\ny$) imply that  $K(\tilde{G}_1(\Xb_1\pr \thetab\n))-K(\tilde{G}_1(\Xb_1\pr \thetab))$ is $o_{\rm P}(1)$ under ${\rm P}_{\thetab;g_1}\n$ as $\ny$. Since $K$ is continuous  on a compact support (and is therefore bounded, see Assumption D), this convergence also holds in quadratic mean. Similarly, the boundedness of $\Sb_{\thetab}(\Xb_1)$ and $\Sb_{\thetab\n}(\Xb_1)$ together with Lemma~\ref{signconver} yields the result of (a). \vspace{2mm}

\noindent We now turn to the proof of \noindent {(b)}. Parts (i), (ii) and (iii) readily state that, under ${\rm P}_{\thetab; g_1}\n$ and for $\ny$,
\begin{equation}\label{equ1a} \Deltab_{{\thetab}; K, g_1}\n \stackrel{\mathcal{L}}{\rightarrow} {\mathcal N}_k \left({\bf 0}, \Gamb_{\thetab; K} \right), \end{equation}
and that under the sequence of local alternatives ${\rm P}_{\thetab\n; g_1}\n$, as $\ny$,
\begin{equation*}\label{equ2} \Deltab_{{\thetab}; K, g_1}\n- \Gamb_{\thetab; K, g_1}{\bf t}\n \stackrel{\mathcal{L}}{\rightarrow} {\mathcal N}_k \left({\bf 0}, \Gamb_{\thetab; K} \right). \end{equation*}
Now, it follows from ULAN that the model is locally and asymptotically linear (see, e.g., \cite{Bic1982}). Therefore, Assumption (D), the continuity of $\Gamb_{\thetab; K, g_1} $ and contiguity entail that, under ${\rm P}_{\thetab; g_1}\n$,
\begin{equation}\label{equ3} \Deltab_{{\thetab}\n; K, g_1}\n+ \Gamb_{\thetab; K, g_1} {\bf t}\n \stackrel{\mathcal{L}}{\rightarrow} {\mathcal N}_k \left({\bf 0}, \Gamb_{{\thetab}; K}\right) \end{equation}
\Black as $\ny$. We have shown in (a) that $ \Deltab_{{\thetab}\n; K, g_1}\n- \Deltab_{{\thetab}; K, g_1}\n-{\rm E}[ \Deltab_{{\thetab}\n; K, g_1}\n]$ is $o_{L^2}(1)$ under ${\rm P}_{\thetab; g_1}\n$ as $\ny$. Therefore, in view of (\ref{equ1a}), we have that
\begin{equation}\label{equ4} \Deltab_{{\thetab}\n; K, g_1}\n-{\rm E}[ \Deltab_{{\thetab}\n; K, g_1}\n] \stackrel{\mathcal{L}}{\rightarrow} {\mathcal N}_k \left({\bf 0},  \Gamb_{{\thetab}; K}\right). \end{equation}
Comparing (\ref{equ3}) and (\ref{equ4}), it follows that ${\rm E}[ \Deltab_{{\thetab}\n; K, g_1}\n]+\Gamb_{\thetab; K,g_1}{\bf t}\n$ is $o(1)$ as $\ny$ under ${\rm P}_{\thetab; g_1}\n$ which is the desired result.

Since we have proved both (a) and (b), the claim of part (iv) of the proposition holds, which concludes the proof. \cqfd \vspace{8mm}

\par

\noindent {\large\bf Acknowledgements}\\

\noindent The research of Christophe Ley is supported by a Mandat de Charg\'e de Recherche from the Fonds National de la Recherche Scientifique, Communaut\' e fran\c caise de Belgique. 
The research of Yvik Swan is supported by a Mandat de Charg\'e de Recherche from the Fonds National de la Recherche Scientifique, Communaut\' e fran\c caise de Belgique.
The authors sincerely thank the editor, associate editor and referees  for their careful reading of the paper which has lead to significant improvements.

\bibliographystyle{elsarticle-harv}

\vskip .85cm
\noindent
Christophe Ley\\
D\'{e}partement de Math\'{e}matique and E.C.A.R.E.S.\\
Universit\'{e} Libre de Bruxelles\\
Boulevard du Triomphe\\
B-1050 Bruxelles\\
E-mail: chrisley@ulb.ac.be\\

\vskip .65cm
\noindent
Yvik Swan\\
D\'{e}partement de Math\'{e}matique\\
Universit\'{e} Libre de Bruxelles\\
Boulevard du Triomphe\\
B-1050 Bruxelles\\
E-mail: yvswan@ulb.ac.be\\

\vskip .65cm
\noindent
Baba Thiam\\
EQUIPPE-GREMARS\\
Universit\' e Lille III\\
Domaine Universitaire du Pont de Bois, BP 60149\\
F-59653 Villeneuve d'Ascq Cedex\\
E-mail: baba.thiam@univ-lille3.fr\\

\vskip .65cm
\noindent
Thomas Verdebout\\
EQUIPPE-GREMARS\\
Universit\' e Lille III\\
Domaine Universitaire du Pont de Bois, BP 60149\\
F-59653 Villeneuve d'Ascq Cedex\\
E-mail: thomas.verdebout@univ-lille3.fr\\

\end{document}